\documentclass{amsart}
\usepackage[utf8]{inputenc}

\usepackage{amsthm}
\usepackage{amssymb}
\usepackage{graphicx}
\usepackage{comment}
\usepackage{amsmath}
\usepackage{breqn}
\makeatletter
\numberwithin{equation}{section}
\numberwithin{figure}{section}
\theoremstyle{plain}

 \theoremstyle{remark}

\newtheorem{theorem}{Theorem}[section]

\newtheorem{lemma}[theorem]{Lemma}

\newtheorem{remark}[theorem]{Remark}

\usepackage{amsthm}
\usepackage{color}
\usepackage[dvipsnames]{xcolor}
\usepackage{subcaption}
\usepackage{enumerate}
\usepackage[draft]{todonotes}
\usepackage[american]{babel}
\usepackage{geometry}

\usepackage{color,soul}

\newcommand{\ri}{\mathrm{i}}
\newcommand{\eig}{\mathrm{eig}}

\newcommand{\tr}{\mathrm{trace}}
\newcommand{\erf}{\mathrm{erf}}
\newcommand{\sign}{\mathrm{sign}}
\newcommand{\erfi}{\mathrm{erfi}}

\def\Id {{ \rm Id}}

\makeatother

\providecommand{\remarkname}{Remark}
\providecommand{\theoremname}{Theorem}

\title{Spectral Analysis and Hydrodynamic Manifolds for the Linearized Shakhov Model}
\author{Florian Kogelbauer}
\author{Ilya Karlin}
\date{\today}

\begin{document}

\maketitle

\begin{abstract}
We perform a complete spectral analysis of the linearized Shakhov model involving two relaxation times $\tau_{\rm fast}$ and $\tau_{\rm slow}$. 
Our results are based on spectral functions derived from the theory of finite-rank perturbations, which allows us to infer the existence of a critical wave number $k_{\rm crit}$ limiting the number of discrete eigenvalues above the essential spectrum together with the existence of a finite-dimensional slow manifold defining non-local hydrodynamics. We discuss the merging of hydrodynamic modes as well as the existence of second sound and the appearance of ghost modes beneath the essential spectrum in dependence of the Prandtl number. 
\end{abstract}

\section{Introduction}

Hydrodynamic closures derived from kinetic theory are a fruitful research direction in statistical physics \cite{gorbankarlin2013,saint2014mathematical} and of basic interest for the investigation of kinetic models \cite{chapman1990mathematical}. The fundamental question: What is the connection between kinetic equations (in the small relaxation regime) and the equations for the motion of continua?  Or, to phrase it differently: Can the governing equations of fluid dynamics be rigorously derived from kinetic theory?\\

This problem has a long history. Famously, in his speech at the International Congress of Mathematics's in Paris in 1900, Hilbert proposed a program to derive the passage from the atomistic view of fluids and gases to the motion of continua \cite{hilbert2000mathematical}. A modern interpretation of this challenge, known as "Hilbert's sixth problem" in this context, aims to proof the convergence of kinetic models, such as the Boltzmann equation, to hydrodynamic models, such as the Navier--Stokes equation \cite{SRay2007,saint2014mathematical}.\\

There are several ways to tackle this problem \cite{truesdell1980fundamentals}. Assuming that the collision term scales as $\varepsilon^{-1}$, a widely used approach is to expand the density function as a formal power series in $\varepsilon$ (the Knudsen number), called Chapman--Enskog series \cite{chapman1990mathematical}. Indeed, the zeroth order PDE obtained from this (singular) Taylor expansion gives the Euler equation, while the first order PDE reproduces the Navier--Stokes equation. We stress, however, that this holds on a formal level only.\\

While the the solution to the underlying kinetic equation decays to the global equilibrium due to increase of entropy, higher-order terms in the Chapman--Enskog expansion might exhibit instabilities. In \cite{bobylev1982chapman}, it was first shown that an expansion in terms of Knudsen number can lead to nonphysical properties of the hydrodynamic models: At order two (Burnett equation \cite{chapman1990mathematical}), the dispersion relation shows a change of sign, thus leading to modes which grow in energy (Bobylev instability). Indeed, as pointed out by, e.g., Slemrod \cite{slemrod2012chapman}, convergence of the singular expansion to the leading-order equation is by no means obvious: the formation of shocks might be an obstacle to global uniform convergence in the sense of solutions \cite{SLEMROD20131497}. Furthermore, the expansion of a non-local operator in frequency space in terms of (local) differential operators might be problematic. Therefore, Rosenau suggested a non-local closure \cite{PhysRevA.40.7193}. A different approach is to sum the Chapman--Enskog series for all orders. This was achieved for the three-component Grad system by Gorban and Karlin in a series of papers \cite{gorbankarlin2013,gorban1994method,gorban2005invariant}.\\

In this work, we approach the problem from the angle of spectral theory. Investigations of the spectrum of linearized kinetic operators date back to Hilbert himself \cite{hilbert1912grundzuge}. Carleman \cite{carleman1957problemes} proved that the essential spectrum remains the same under a compact perturbation (Weyl's theorem) in the hard sphere case and was able to estimate the spectral gap. This result was generalized to a broader class of collision kernels by Grad \cite{grad1963asymptotic} and to soft potentials in \cite{caflisch1980boltzmann}. For spatially uniform Maxwell molecules, a complete spectral description was derived in \cite{Bobylev1988} (together with exact special solutions and normal form calculations for the full, nonlinear problem), see also \cite{chang1970studies}. In \cite{ellis1975first}, some fundamental properties of the spectrum of a comparably broad class of kinetic operators was derived. In particular, the existence of eigenvalue branches and asymptotic expansion of the (small) eigenvalues for vanishing wave number was derived. The analysis carried out in \cite{ellis1975first}, however, does not extend to large wave numbers or to properties of the discrete spectrum close to the essential spectrum (accumulation of eigenvalues). For convergence results to the Euler and Navier--Stokes equation based on spectral insights, we refer to the classical papers \cite{bardos1991classical,nishida1978fluid}\\

This paper is devoted to the spectral analysis of the Shakhov model \cite{shakhov1968generalization} linaerized around a global Maxwellian, see also \cite{BAE2023103742}. While the BGK equation only has one global relaxation time $\tau$, the Shakhov model has two different time scales $\tau_{\rm fast}$ and $\tau_{\rm slow}$, which, in particular, allows to consider a family of kinetic equations with varying Prandtl number. While the BGK equation has Prandtl number one, the Shakhov model admits Prandtl numbers $0<\rm Pr<1$ (and also the somewhat nonphysical $\rm Pr>1$, see Section \ref{spectral5}) thus give a more realistic approximation of relaxation times. In particular, it allows for the Prandtl number $\rm Pr={2}/{3}$, see, e.g. \cite{chapman1990mathematical}. \\

In our previous papers, similar considerations were already carried out for the three-component Grad system \cite{Kogelbauer2019}, the one-dimensional BGK equation with mass conservation only \cite{kogelbauer2021}, and recently for the three-dimensional BGK equation with the mass, momentum and energy conservation \cite{kogelbauer2023exact}. 
Similar considerations have been carried out in \cite{ThomasCarty2017,carty2017elementary} for the one-dimensional linear BGK with one conservation law that of mass in the context of grossly determined solutions (in the sense of \cite{truesdell1980fundamentals}). In \cite{achleitner2017multi}, (optimal) decay rates for various simplified BGK models where derived in the context of hypocoercivity \cite{villani2006hypocoercivity}.\\

In comparison with the previously mentioned kinetic models, Shakhov's equation admits more realistic behavior, i.e., mimics properties of the full Boltzmann equation more closely. Indeed, due to the presence of two time-scales $(\tau_{\rm fast},\tau_{\rm slow})$ in Shakhov's model does not only admit hydrodynamic modes, but non-hydrodynamic (fast) modes as well. In difference with the BGK equation, whose modal branches stay coherent until they mix with the essential spectrum \cite{kogelbauer2023exact}, modal branches of the Shakhov equation can mix and produce new branches. For instance, two diffusion modes can collide and produce a second pair of acoustic modes, see Section \ref{spectral4}.  
Moreover, Shakhov's model and its versions are widely used in gas-dynamics applications, in particular, in the lattice Boltzmann computations of compressible flows, see, e.g., \cite{frapolli2015entropic}. We emphasize that the techniques outlined in this paper can easily be applied to a wide class of similar kinetic models, such as the ES-BGK \cite{holway1965kinetic}.

In the following, we will give a complete and (up to a solution of a transcendental equation) explicit description of the spectrum of the Shakhov model linearized around a global Maxwellian. We will show the existence of \textit{finitely many} discrete eigenvalues above the essential spectrum as well as the existence of a critical wave number for each family of modes. More precisely, we prove the following:
\begin{theorem}
The spectrum of the non-dimensional linearized Shakhov operator $\mathcal{L}$ with relaxation times $\tau_{\rm fast}$ and $\tau_{\rm slow}$ around a global Maxwellian is given by
\begin{equation}
\sigma(\mathcal{L}) = \left\{-\frac{1}{\tau_{\rm fast}}+\ri\mathbb{R}\right\}\cup\bigcup_{|\mathbf{k}|<k_{\rm crit}}\bigcup_{N\in \text{Modes}(|\mathbf{k}|,\rm Pr)}\{\lambda_{N}(\mathbf{k}|)\},
\end{equation}
where $\text{Modes}$ denotes the set of modes (branches), which might change with wave number and Prandtl number. This is due to a collision of roots and a subsequent bifurcation (for details, we refer to Section \ref{spectral4}).
The essential spectrum is given by the line $\Re\lambda=-\frac{1}{\tau_{\rm \rm fast}}$, while the discrete spectrum consists of a \textit{finite} number of discrete, isolated eigenvalues. Along with each family of modes (up to merging), there exists a critical wave number $k_{crit,N}$, limiting the range of wave numbers for which $\lambda_N$ exists. For small wave numbers and $\rm Pr$ close to one, the set of modes is given by 
 \begin{equation}\label{mode1thm}
    \rm Modes_1 = \{\rm shear_1,diff_1,ac_1,ac_1*,shear_2,diff_2\},
 \end{equation}
 while the higher wave-number and $\rm Pr$ closer to zero, set of modes is given by
  \begin{equation}\label{mode2thm}
     \rm Modes_2 = \{\rm shear_1,ac_1,ac_1*,shear_2,ac_2,ac_2*\},
 \end{equation}
 where $\lambda_{\rm shear,1},\lambda_{\rm shear, 2}$ denote the (real) primary and secondary shear modes (double degenerated), $\{\lambda_{\rm ac,1}, \lambda_{\rm ac,1}^*\}, \{\lambda_{\rm ac,2}, \lambda_{\rm ac,2}^*\}$ denote pairs of complex conjugated roots, the primary and secondary acoustic modes and the real roots $\lambda_{\rm diff,1}, \lambda_{\rm diff,2}$ denote the (real) primary and secondary diffusion modes, respectively. 
\end{theorem}

For discussion of different branches of eigenvalues in kinetic models, we refer to \cite{ellis1975first,ford1988dispersion}. 
Our proof is based on the theory of finite-rank perturbations (see, e.g., \cite{WEINSTEIN1974604}), together with some properties of the plasma dispersion function, collected in the Appendix for the sake of completeness. Furthermore, we give a hydrodynamic interpretation of the results by considering the dynamics on the slow hydrodynamic manifold (linear combination of eigenspaces).\\

The paper is structured as follows: In Section \ref{definitions}, we introduce some notation and give some basic definitions. In Section \ref{Prelim}, we formulate the fundamental equations (the detailed linearization around a global Maxwellian as well as the non-dimensionalization are included to the Appendix). Section \ref{spectralan} is devoted to the spectral analysis of the linear part, including the derivation of a spectral function describing the discrete spectrum completely. We also give a proof of the finiteness of the hydrodynamic spectrum together with a description of the modes (primary shear, primary diffusion, primary acoustic, secondary shear and secondary diffusion) in frequency space. We also comment on the merging of branches and the formation of a secondary acoustic branch, called \textit{second sound}  
for a certain range of Prandtl numbers, see, e.g., \cite{peshkov2013second} or \cite{hardy1970phonon} for an analogous phenomenon in solids (phonons).\\
Finally, in Section \ref{sechydro}, we write down the hydrodynamic manifold as a linear combination of eigenvectors and derive a closed system for the linear hydrodynamic variables.\\

\section{Notation and Basic Definitions}\label{definitions}
For a wave vector $\mathbf{k}\in\mathbb{Z}^3$, $\mathbf{k}=(k_1,k_2,k_3)$, we denote its wave number as 
\begin{equation}
k:=|\mathbf{k}|=\sqrt{k_1^2+k_2^2+k_3^2}.
\end{equation}
Let $\mathcal{H}$ denote a Hilbert space and let $\mathbf{T}:\mathcal{H}\to\mathcal{H}$ be a linear operator with domain of definition $\mathcal{D}(\mathcal{H})$. We denote the spectrum of $\mathbf{T}$ as $\sigma(\mathbf{T})$ and its resolvent set as $\rho(\mathbf{T})$.\\
The spectral analysis of the main operator $\mathcal{L}$ of the paper (to be defined later) will be carried out on the Hilbert space 
\begin{equation}
\mathcal{H}_{\mathbf{x},\mathbf{v}}=L^2_{\mathbf{x}}(\mathbb{T}^3) \times L^2_{\mathbf{v}}(\mathbb{R}^3,e^{-\frac{|\mathbf{v}|^2}{2}}),
\end{equation}
together with the inner product 
\begin{equation}
\langle f, g \rangle_{\mathbf{x},\mathbf{v}} = \frac{1}{(2\pi)^{3+\frac{3}{2}}} \int_{\mathbb{T}^3}\int_{\mathbb{R}^3} f(\mathbf{x},\mathbf{v}) g^*(x,\mathbf{v})\,  e^{-\frac{|\mathbf{v}|^2}{2}} d\mathbf{v} d\mathbf{x},
\end{equation}
where the star denotes complex conjugation. Because of the unitary properties of the Fourier expansion, we can slice the space $\mathcal{H}$ for each wave number $\mathbf{k}$ and analyze the operator $\mathcal{L}_{\mathbf{k}}$ (restriction of $\mathcal{L}$ to the wave number $\mathbf{k}$) on the Hilbert space
\begin{equation}
\mathcal{H}_{\mathbf{v}} = L^2_{\mathbf{v}}(\mathbb{R}^3,e^{-\frac{|\mathbf{v}|^2}{2}}),
\end{equation}
together with the inner product
\begin{equation}\label{innerprod}
\langle f, g \rangle_{\mathbf{v}} :=\frac{1}{(2\pi)^{\frac{3}{2}}}\int_{\mathbb{R}^3} f(\mathbf{v})g^*(\mathbf{v}) e^{-\frac{|\mathbf{v}|}{2}}d\mathbf{v}.
\end{equation}


For $\mathbf{n}=(n_1,n_2,n_3)$, the three-dimensional (multidimensional) Hermite polynomials $\{H_{\mathbf{n}}\}_{\mathbf{n}\in\mathbb{N}^3}$ are defined via the generating function
\begin{equation}\label{Hermitegen}
e^{\mathbf{a}\cdot\mathbf{v}-\frac{|\mathbf{a}|^2}{2}} = \sum_{|\mathbf{n}|=0} c_{\mathbf{n}}(\mathbf{a}) H_{\mathbf{n}}(\mathbf{v}),
\end{equation}
for the coefficients 
\begin{equation}
    c_{\mathbf{n}}(\mathbf{a})=\frac{\mathbf{a}^{\mathbf{n}}}{\mathbf{n}!}. 
\end{equation}
Let us denote the $j$-th standard basis vector of $\mathbb{R}^3$ as $\mathbf{e}_j$. The three-dimensional Hermite polynomials obey the recurrence relation
\begin{equation}
 H_{\mathbf{n}+\mathbf{e}_j}(\mathbf{v}) = v_j H_{\mathbf{n}}(\mathbf{v})- n_j H_{\mathbf{n}-\mathbf{e}_j}(\mathbf{v})   ,
\end{equation}
for $j=1,2,3$,
as well as the orthogonality relation
\begin{equation}
    \frac{1}{(2\pi)^{\frac{3}{2}}}\int_{\mathbb{R}^3}H_{\mathbf{m}}(\mathbf{v})H_{\mathbf{n}}(\mathbf{v}) e^{-\frac{|\mathbf{v}|^2}{2}}\, d\mathbf{v} = \frac{1}{\mathbf{m}!} \delta_{\mathbf{m},\mathbf{n}},
\end{equation}
where $\delta_{\mathbf{m},\mathbf{n}}$ is the Kronecker delta. In particular, the sequence of one-dimensional Hermite polynomials $H_n(v)$ obey the recurrence relation
\begin{equation}\label{recHermite}
    H_{n+1}(v) = v H_{n}(v) - n H_{n-1}(v),
\end{equation}
which, by \eqref{Hermite}, implies the differential recurrence relation
\begin{equation}\label{recHermite2}
    H_{n+1}(v) = H_n'(v)-vH_n(v).  
\end{equation}
We introduce the \textit{plasma dispersion function} as the integral 
\begin{equation}\label{defZ}
    Z(\zeta) = \frac{1}{\sqrt{2\pi}} \int_{\mathbb{R}} \frac{e^{-\frac{v^2}{2}}}{v-\zeta}\, dv,
\end{equation}
for any $\zeta\in\mathbb{C}\setminus\mathbb{R}$. The function $Z$ is analytic on each half plane $\{\Im(\zeta)>0\}$ and $\{\Im(\zeta)<0\}$ and satisfies the complex differential equation
\begin{equation}\label{diffZ}
    \frac{dZ}{d\zeta} = -\zeta Z-1.
\end{equation}
We collect further useful properties of $Z$ in the Appendix. 
\begin{remark}
The plasma dispersion function \eqref{defZ} - as its name suggests - appears in plasma physics in the context of Landau damping \cite{fitzpatrick2014plasma}. An detailed description of this widely used non-elementary function is presented in \cite{fried2015plasma}. 
\end{remark}

\section{Preliminaries, Linearization and Non-Dimensionalization}\label{Prelim}
Consider the kinetic model 
\begin{equation}\label{maineq}
\frac{\partial F}{\partial t} +\mathbf{v}\cdot\nabla F = -\frac{1}{\tau_{\rm fast}}Q_{fs}(F),
\end{equation}
for an unknown distribution function $F$ and the collision operator
\begin{equation}
Q_{fs}(F)=F-F^{eq}(n[F],\mathbf{u}[F],T[F])\left(1+(1-Pr)\frac{\mathbf{q}[F]\cdot(\mathbf{v}-\mathbf{u}[F])}{Rp[F]T[F]}\left(\frac{|\mathbf{v}-\mathbf{u}[F]|^2}{5RT[F]}-1\right)\right).
\end{equation}
Here, $R$ denotes the gas constant, while 
\begin{equation}
\rm Pr = \frac{\tau_{\rm fast}}{\tau_{\rm slow}}\leq 1,
\end{equation}
denotes the Prandtl number for the fast time scale $\tau_{\rm fast}$ and the slow time scale $\tau_{\rm slow}$. The physical units are given as $[R]=m^2kg s^{-2}K^{-1}$ and $[RT] = m^2 kg s^{-2}$ respectively, while the Prandtl number is dimensionless. The number density $n$, the velocity $\mathbf{u}$, the pressure $p$ and the heat flux $\mathbf{q}$ are defined by
\begin{equation}\label{defmacro}
\begin{split}
n[F]&=\int_{\mathbb{R}^3}F\, d\mathbf{b},\\
\mathbf{u}[F]&=\frac{1}{n[F]}\int_{\mathbb{R}^3}\mathbf{v}F\, d\mathbf{b},\\
p[F]&=\frac{m}{3}\int_{\mathbb{R}^3}|\mathbf{v}-\mathbf{u}[F]|^2F\, d\mathbf{v},\\
\mathbf{q}[F] & = \frac{m}{2}\int_{\mathbb{R}^3} (\mathbf{v}-\mathbf{u}[F])|\mathbf{u}[F]-\mathbf{v}|^2 F\, d\mathbf{v},
\end{split}
\end{equation}
while the temperature $T$ is defined through the relation
\begin{equation}
T[F]=\frac{p[T]}{mRn[T]}.
\end{equation}
We stress that the macroscopic quantities $(n,\mathbf{u},T,\mathbf{q})$ depend on $F$ through the functional relationship \eqref{defmacro}, but are independent of the velocity $\mathbf{v}$.\\
Equation \eqref{maineq} is called \textit{Shakhov's S-model} and was first derived in \cite{shakhov1968generalization} as a generalization to the BGK equation, allowing for two time scales, which, in particular, allows to define a family of models with varying Prandtl number. To ease notation in the following calculations, we set
\begin{equation}
    \tau = \tau_{\rm fast}
\end{equation} 
and define the dimensionless parameter
\begin{equation}
r:=1-\rm Pr,
\end{equation}
which satisfies $0\leq r\leq 1$ as well.
\begin{remark}
    The Prandtl number (and the parameter $r$) allow us to define an interpolation between the three-dimensional BGK equation ($\rm Pr = 1$ r $r=0$), see e.g. \cite{wang_wu_ho_li_li_zhang_2020}, and a model with maximal separation between fast and slow time-scale ($\rm Pr = 0$ or $r=1$). The different properties of the spectra, see Section \ref{spectralan}, vary for different values of $r$. 
\end{remark}

For $n\geq 0$, we define the linear moments
\begin{equation}\label{defmoment}
\mathbf{M}_n(\mathbf{x},t) = \int_{\mathbb{R}^3} F(\mathbf{x},\mathbf{v},t)\mathbf{v}^{\otimes n}\, d\mathbf{v},
\end{equation}
where $\mathbf{v}^{\otimes 0}=1$, $\mathbf{v}^{\otimes 1}=\mathbf{v}$ and 
\begin{equation}
\mathbf{v}^{\otimes n}=\underbrace{\mathbf{v}\otimes...\otimes \mathbf{v}}_{n-\text{times}},
\end{equation}
for $n\geq 2$ is the $n^{th}$ tensor power.
For compactness in the presentation, we also define the special vector
\begin{equation}
\tilde{\mathbf{M}}_3 = \int_{\mathbb{R}^3} F \mathbf{v}|\mathbf{v}|^2\, d\mathbf{v}. 
\end{equation}
The hydrodynamic variables $(n,\mathbf{u},T,\mathbf{q})$ are related to the moments \eqref{defmoment} via
\begin{equation}
    \begin{split}
      \mathbf{M}_0  & = n,\\
      \mathbf{M}_1 & = n\mathbf{u},\\
      \tr\mathbf{M}_2 & = n\left(|\mathbf{u}|^2+3RT\right)=n|\mathbf{u}|^2+3\frac{p}{m},\\
      |\mathbf{u}|^2\mathbf{u}-2\mathbf{M}_2\mathbf{u}+\tilde{\mathbf{M}}_3  & = \frac{2}{m}\mathbf{q}+\frac{3p}{2}\mathbf{u}, 
    \end{split}
\end{equation}
which can be inverted to
\begin{equation}
    \begin{split}
        n & = \mathbf{M}_0 ,\\
      \mathbf{u} & = \frac{\mathbf{M}_1}{\mathbf{M}_0} ,\\
      p & = \frac{m}{3} \tr\mathbf{M}_2-\frac{m}{3}\frac{|\mathbf{M}_1|^2}{\mathbf{M}_0},\\
      \mathbf{q} & = -\frac{3}{2}\left(\frac{m}{3} \tr\mathbf{M}_2-\frac{m}{3}\frac{|\mathbf{M}_1|^2}{\mathbf{M}_0}\right)\frac{\mathbf{M}_1}{\mathbf{M}_0}+\frac{m}{2}\tilde{\mathbf{M}}_3+\frac{m}{2}|\mathbf{u}|^2\mathbf{u}-m\mathbf{M}_2\mathbf{u}.
    \end{split}
\end{equation}

Equation \eqref{maineq} admits a global Maxwellian,
\begin{equation}\label{Maxwellian}
F^{eq}_0(\mathbf{v})=\frac{n_0}{(2\pi R T_0)^{\frac{3}{2}}}e^{-\frac{|\mathbf{v}|^2}{2RT_0}},
\end{equation}
as an equilibrium solution. Here, the equilibrium number density $n_0$ and the equilibrium temperature $T_0$ are constants. In the following, we will be interested in the dynamics of \eqref{maineq} close to the stationary solution \eqref{Maxwellian}, i.e., the linearized dynamics of \eqref{maineq} around  \eqref{Maxwellian}.\\
The Shakhov model linearized around \eqref{Maxwellian} in non-dimensional form is given by
\begin{equation}\label{linmain}
    \begin{split}
        \frac{\partial f}{\partial t} =-\mathbf{v}\cdot\nabla_{\mathbf{x}}f-\frac{1}{\tau}f+\frac{1}{\tau} (2\pi)^{-3/2}e^{-\frac{|\mathbf{v}|^2}{2}}\left[\left(\frac{5-|\mathbf{v}|^2}{2}\right)\mathbf{m}_0+\left(1+\frac{r}{2}\left(\frac{|\mathbf{v}|^2-5}{5}\right)\right)\mathbf{v}\cdot\mathbf{m}_1\right.\\
\left.+\left(\frac{|\mathbf{v}|^2-3}{2}\right)\tr\mathbf{m}_2+r\left(\frac{|\mathbf{v}|^2-5}{10}\right)\mathbf{v}\cdot\tilde{\mathbf{m}}_3\right].
    \end{split}
\end{equation}
For the details, we refer to Appendix I. Equation \eqref{linmain} will serve as the basis for the spectral analysis performed in the following section.

\section{Spectral Analysis of the Linearized Two-Timescale Operator}\label{spectralan}

In this section, we will carry out a complete spectral analysis of the right-hand side of \eqref{linmain}, following the approach in \cite{kogelbauer2023exact}. This will allow us to draw conclusions on the decay properties of hydrodynamic variables, the existence of a critical wave number and the hydrodynamic closure. After reformulating the problem in frequency space, we will use the resolvent calculus to formulate a condition for the discrete spectrum (Subsection \ref{spectral1}). Then, we will use properties of the plasma dispersion function (see Appendix) to define a spectral function $\Sigma_{|\mathbf{k}|,\tau}$, whose zeros coincide with the discrete, isolated eigenvalues (Subsection \ref{spectral2}). These families of eigenvalues are described in more detail in Subsection \ref{spectral3}. Then, in Subsection \ref{spectral3b}, we prove the existence of a critical wave number $k_{\rm crit}$ such that $\Sigma_{|\mathbf{k}|,\tau}$ has no zeros (i.e., there exists no eigenvalues) for $|\mathbf{k}|>k_{\rm crit}$. In Subsection \ref{spectral4}, we take a closer look at the branches of eigenvalues (modes) and the merging of diffusive branches (second sound). Finally, in Subsection \ref{spectral5}, we discuss the existence of eigenvalues below the essential spectrum (ghost modes) for $r<0$ ($\rm Pr>1$). \\

\subsection{Description of the discrete spectrum}\label{spectral1}
In the following, we rescale the density $f$ with a global, non-dimensional Maxwellian,
\begin{equation}\label{rescale}
    f\mapsto (2\pi)^{-3/2}e^{-\frac{|\mathbf{v}|^2}{2}} f,
\end{equation}
which allows us to divide by the Gaussian in \eqref{linmain} and interpret the moments \eqref{defm} as projections relative to the inner product \eqref{innerprod}. We define the following set of basis functions
\begin{align}\label{basis}
e_0(\mathbf{v}) &= 1,\qquad e_4(\mathbf{v}) = \frac{|\mathbf{v}|^2-3}{\sqrt{6}},\\
e_1(\mathbf{v}) &= v_1,\qquad e_5(\mathbf{v}) = v_1\frac{|\mathbf{v}|^2-5}{\sqrt{10}},\\
e_2(\mathbf{v}) &= v_1, \qquad e_6(\mathbf{v})  = v_2\frac{|\mathbf{v}|^2-5}{\sqrt{10}},\\
e_3(\mathbf{v}) &= v_3, \qquad e_7(\mathbf{v}) = v_3\frac{|\mathbf{v}|^2-5}{\sqrt{10}},
\end{align}
which satisfy the orthonormality condition
\begin{equation}
\langle e_n, e_m \rangle_{\mathbf{v}} = \delta_{nm}, \quad \text{for} \quad  0\leq n,m \leq 7.
\end{equation}
Defining
\begin{equation}\label{fcoef}
f_j=\langle e_j,f\rangle_{\mathbf{v}},
\end{equation}
we can infer the following relations between the moments and the coefficients \eqref{fcoef}:
\begin{equation}
\begin{split}
\frac{5-|\mathbf{v}|^2}{2}\mathbf{m}_0&=\frac{5-|\mathbf{v}|^2}{2}f_0=f_0e_0-\frac{\sqrt{6}}{2}f_0e_4,\\
\left(\frac{r}{2}\left(\frac{|\mathbf{v}|^2-5}{5}\right)+1\right)\mathbf{v}\cdot\mathbf{m}_1 &=f_1e_1+f_2e_2+f_3e_3-\frac{r\sqrt{10}}{2}(f_1e_5+f_2e_6+f_3e_7),\\
\frac{|\mathbf{v}|^2-3}{6}\tr\mathbf{m}_2&= e_4 \frac{1}{\sqrt{6}} \int_{\mathbb{R}} f |\mathbf{v}|^2\, d\mathbf{v}=e_4 \frac{1}{\sqrt{6}}\left( \int_{\mathbb{R}} f (|\mathbf{v}|^2-3)\, d\mathbf{v}+3\mathbf{m}_0\right)=f_2e_4+\frac{3}{\sqrt{6}}f_0e_4\\
r\left(\frac{|\mathbf{v}|^2-5}{10}\right)\mathbf{v}\cdot\tilde{\mathbf{m}}_3 & = \frac{r}{\sqrt{10}}(e_5,e_6,e_7)\cdot \int_{\mathbf{R}^3} f\mathbf{v}|\mathbf{v}|^2\, d\mathbf{v} = (e_5,e_6,e_7)\cdot \int_{\mathbf{R}^3} \Big[f(e_5,e_6,e_7)+\frac{5r}{\sqrt{10}}\mathbf{v}f\Big]\, d\mathbf{v}\\
& = r(f_5e_5+f_6e_6+f_7e_7)+\frac{5r}{\sqrt{10}}(f_1e_5+f_2e_6+f_3e_7)
\end{split}
\end{equation}
We bundle the basis functions \eqref{basis} into a vector 
\begin{equation}
\mathbf{e}=\{e_j\}_{0\leq j\leq 7},
\end{equation}
and define the matrix
\begin{equation}
\mathbf{D}_r=\text{diag}(1,1,1,1,1,r,r,r),
\end{equation}
together with the following 
\begin{equation}
\mathbb{B}_{8,r}=(\mathbb{P}_5+r\mathbb{P}_8)f=\langle f, \mathbf{e}\rangle_{\mathbf{v}} \cdot \mathbf{D}_r\mathbf{e},
\end{equation}
as the scaled sum of two finite-rank projections
\begin{equation}
\mathbb{P}_5f = \sum_{n=0}^4 \langle f,e_n\rangle e_n,\qquad  \mathbb{P}_8f = \sum_{n=5}^8 \langle f,e_n\rangle e_n.
\end{equation}
The linear operator appearing as the right-hand side of equation \eqref{linmain} (together with the rescaling \eqref{rescale}) then takes the simple form 
\begin{equation}\label{defL}
\mathcal{L}=-\mathbf{v}\cdot\nabla_{\mathbf{x}}-\frac{1}{\tau}+\frac{1}{\tau}\mathbb{B}_{8,r},
\end{equation}
and equation \eqref{linmain} becomes
\begin{equation}\label{eqmainlinear}
    \frac{\partial f}{\partial t} = \mathcal{L}f.
\end{equation}
\begin{remark}
Let us recall that any function $f\in \mathcal{H}_{\mathbf{v}}$ admits a unique expansion as a multi-dimensional \textit{Hermite series}:
\begin{equation}\label{Hermite}
f(\mathbf{v})=\sum_{n=0}^{\infty} \mathbf{f}_n:\mathbf{H}_n(\mathbf{v}),
\end{equation}
where $\mathbf{H}_n$ is defined in \eqref{Hermitegen} and $\mathbf{f}_n$ is an $n$-tensor. Since the eight basis vectors \eqref{basis} appear in the expansion \eqref{Hermite} via an orthogonal splitting, we have that
\begin{equation}\label{ortho}
\langle \mathbb{P}_5f, (1-\mathbb{P}_{5})f\rangle_{\mathbf{v}}=0,\qquad \langle \mathbb{P}_8f, (1-\mathbb{P}_{8})f\rangle_{\mathbf{v}}=0
\end{equation}
 for any $f\in\mathcal{H}_{\mathbf{v}}$. Hermite expansions were famously used by Grad in his seminal paper \cite{grad1949kinetic} to establish finite-moment closures.
\end{remark}
From 
\begin{equation}\label{dissipative}
\begin{split}
\langle \mathcal{L}f,f\rangle_{\mathbf{x},\mathbf{v}} & =\langle -\mathbf{v}\cdot\nabla_{\mathbf{x}}f - \frac{1}{\tau}f+\frac{1}{\tau}\mathbb{B}_{8,r}f,f\rangle_{\mathbf{x},\mathbf{v}}\\
&=\int_{\mathbb{T}^3}\int_{\mathbb{R}^3}(-\mathbf{v}\cdot\nabla_{\mathbf{x}}f - \frac{1}{\tau}f+\frac{1}{\tau}(\mathbb{P}_5+r\mathbb{P}_8)f) f e^{-\frac{|\mathbf{v}|^2}{2}}\,d\mathbf{x}d\mathbf{v}\\
&=-\frac{1}{\tau}\int_{\mathbb{T}^3}\int_{\mathbb{R}^3}\Big[f-\mathbb{P}_5f-r\mathbb{P}_8f\Big]f e^{-\frac{|\mathbf{v}|^2}{2}}\,d\mathbf{x}d\mathbf{v}\\
&=-\frac{1}{\tau}\int_{\mathbb{T}^3}\int_{\mathbb{R}^3}\Big[(1-\mathbb{P}_5-\mathbb{P}_8)f+(1-r)\mathbb{P}_8f\Big]f e^{-\frac{|\mathbf{v}|^2}{2}}\,d\mathbf{x}d\mathbf{v}\\
& =-\frac{1}{\tau}\|(1-\mathbb{P}_5-\mathbb{P}_8)f\|^2_{\mathbf{x},\mathbf{v}} - \frac{(1-r)}{\tau} \|\mathbb{P}_8f\|^2_{\mathbf{x},\mathbf{v}},
\end{split}
\end{equation}
where we have assumed that $f$ is sufficiently regular to justify the application of the divergence theorem in $\mathbf{x}$ in order to remove the gradient term as well as \eqref{ortho}. For $0<r<1$ (or $0<\rm Pr<1$). it follows that the operator $\mathcal{L}$ is dissipative and that 
\begin{equation}
\Re\sigma(\mathcal{L})\leq 0.
\end{equation}
On the other hand, for $r>0$, since $\mathbb{P}_5$ and $\mathbb{P}_8$ are orthogonal projections, it follows that
\begin{equation}\label{lowerbound}
\begin{split}
    \langle \mathcal{L}f,f\rangle_{\mathbf{x},\mathbf{v}} & =-\frac{1}{\tau}\int_{\mathbb{T}^3}\int_{\mathbb{R}^3}\Big[f-\mathbb{P}_5f-r\mathbb{P}_8f\Big]f e^{-\frac{|\mathbf{v}|^2}{2}}\,d\mathbf{x}d\mathbf{v}\\
    & = -\frac{1}{\tau}(\|f\|^2_{\mathbf{x},\mathbf{v}}-\|\mathbb{P}_5f\|^2_{\mathbf{x},\mathbf{v}}-r\|\mathbb{P}_8f\|^2_{\mathbf{x},\mathbf{v}})\\
        & \geq -\frac{1}{\tau}\|f\|^2_{\mathbf{x},\mathbf{v}}\\
    \end{split}
\end{equation}
This shows that any solution to \eqref{eqmainlinear} has to converge to zero, i.e., the global Maxwellian is a stable equilibrium up to the conserved quantities from the projected modes.  On the other hand, we infer that the overall convergence rate to equilibrium can be at most $-\frac{1}{\tau}$ for $r>0$. For $r<0$, we can estimate
\begin{equation}\label{lowerboundnegative}
    \begin{split}
        \langle \mathcal{L}f,f\rangle_{\mathbf{x},\mathbf{v}} & =-\frac{1}{\tau}\int_{\mathbb{T}^3}\int_{\mathbb{R}^3}\Big[f-\mathbb{P}_5f-r\mathbb{P}_8f\Big]f e^{-\frac{|\mathbf{v}|^2}{2}}\,d\mathbf{x}d\mathbf{v}\\
    & = -\frac{1}{\tau}(\|f\|^2_{\mathbf{x},\mathbf{v}}-\|\mathbb{P}_5f\|^2_{\mathbf{x},\mathbf{v}}-r\|\mathbb{P}_8f\|^2_{\mathbf{x},\mathbf{v}})\\
        & \geq -\frac{1}{\tau}(\|f\|^2_{\mathbf{x},\mathbf{v}}-r\|\mathbb{P}_8f\|^2_{\mathbf{x},\mathbf{v}})\\
         & \geq \frac{r-1}{\tau}\|f\|^2_{\mathbf{x},\mathbf{v}},
    \end{split}
\end{equation}
where we have used that $\|\mathbb{P}_8f\|^2\leq \|f\|^2$ as well. Consequently, for $r<0$, we can only infer the weaker decay rate  $\frac{r-1}{\tau}$ as opposed to $\frac{1}{\tau}$ for $r>0$. 
\begin{remark}
The weaker decay rate in \eqref{lowerboundnegative} for negative $r$, i.e., Prandtl number larger than one, is already indicative for the existence of parts of the spectrum located below the essential spectrum ($\{\Re\lambda=-\frac{1}{\tau}\}$). Indeed, we will show in Section \ref{spectral5} that for $r<0$, there exist eigenvalues below the essential spectrum for a certain range of wave numbers. 
\end{remark}
Let us proceed with the spectral analysis by switching to frequency space. Since $\mathbf{x}\in\mathbb{T}^3$, we can decompose $f$ in a Fourier series as
\begin{equation}
f(\mathbf{x},\mathbf{v})= \sum_{|\mathbf{k}|=0}^{\infty}\hat{f}(\mathbf{k},\mathbf{v}) e^{\ri \mathbf{x}\cdot\mathbf{k}}, 
\end{equation}
for the Fourier coefficients
\begin{equation}
\hat{f}(\mathbf{k},\mathbf{v}) = \frac{1}{(2\pi)^3}\int_{\mathbb{R}^3} f(\mathbf{x},\mathbf{v})e^{-\ri \mathbf{x}\cdot\mathbf{k}}\, d\mathbf{x}.
\end{equation}
In frequency space, \eqref{defL} becomes
\begin{equation}\label{defLhat}
\hat{\mathcal{L}}_{\mathbf{k}}=-\ri\mathbf{k}\cdot\mathbf{v}-\frac{1}{\tau}+\frac{1}{\tau}\mathbb{B}_{8,r},
\end{equation}
and the spectrum of $\mathcal{L}$ can be calculated from the corresponding operator at each wave vector $\mathbf{k}$:
\begin{equation}
    \sigma(\mathcal{L}) = \bigcup_{\mathbf{k}\in \mathbb{Z}^3} \sigma(\hat{\mathcal{L}}_{\mathbf{k}}). 
\end{equation}
For $\mathbf{k}=0$, we can read off the spectrum of \eqref{defLhat} explicitly. Indeed, since $\hat{\mathcal{L}}_0 = -\frac{1}{\tau}(1-\mathbb{B}_{8,r})$,
we find that
\begin{equation}\label{Lk0}
    \hat{\mathcal{L}}_0 e_j = -\frac{1}{\tau}(1-\mathbb{B}_{8,r})e_j =0,
\end{equation}
for $0\leq j \leq 4$, while
\begin{equation}
    \hat{\mathcal{L}}_0 e_l = -\frac{1}{\tau}(1-\mathbb{B}_{8,r})e_l=-\frac{1}{\tau}(1-r)e_l = -\frac{1}{\tau_{\rm slow}}e_l,
\end{equation}
for $5\leq l \leq 7$. On the other hand, we see from \eqref{Lk0} that $\hat{\mathcal{L}}_0$ just acts like $-\frac{1}{\tau}$ on the orthogonal complement of $\text{span}\{e_j\}_{0\leq j \leq 7}$. This shows that 
\begin{equation}\label{sigmaL0}
    \sigma(\hat{\mathcal{L}}_0) = \left\{-\frac{1}{\tau_{\rm fast}},-\frac{1}{\tau_{\rm slow}},0\right\},
\end{equation}
and the dimensions of the corresponding eigenspaces are given by
\begin{equation}
    \text{codim }\eig\left(-\frac{1}{\tau_{\rm fast}}\right) = 8,\quad \dim \eig\left(-\frac{1}{\tau_{\rm slow}}\right) = 3,\quad \eig\left(0\right) = 5.
\end{equation}
Now, for the following, let us assume that $\mathbf{k}\neq 0$. For more compact calculation, we define the operator
\begin{equation}
    \mathcal{S}_{\mathbf{k}}f = \mathbf{v}\cdot\mathbf{k} f.
\end{equation}
In the calculation of the discrete spectrum of the operator $\hat{\mathcal{L}}_{\mathbf{k}}$, based on the second resolvent identity and finite-rank perturbations, we also follow the presentation in \cite{kogelbauer2023exact} closely. The spectrum of $\hat{\mathcal{L}}_{\mathbf{k}}$ is then given by
\begin{equation}
\begin{split}
\sigma(\hat{\mathcal{L}}_{\mathbf{k}})&= -\frac{1}{\tau} - \sigma\left(\ri \mathcal{S}_{\mathbf{k}}-\frac{1}{\tau}\mathbb{B}_{8,r}\right)\\
&=-\frac{1}{\tau} - \frac{1}{\tau}\sigma\left(\ri \tau\mathcal{S}_{\mathbf{k}}-\mathbb{B}_{8,r}\right).
\end{split}
\end{equation}
Since $\mathbb{B}_{8,r}$ has finite rank, the operator $\ri \tau\mathcal{S}_{\mathbf{k}}-\mathbb{B}_{8,r}$ is compact a compact perturbation of $\ri \tau\mathcal{S}_{\mathbf{k}}$ and we conclude that
\begin{equation}
    \sigma_{ess}(\hat{\mathcal{L}}_{\mathbf{k}}) = -\frac{1}{\tau} - \sigma_{ess}\left(\ri \mathcal{S}_{\mathbf{k}}\right) = -\frac{1}{\tau} +\ri\mathbb{R},
\end{equation}
where we have used that $\sigma(\mathcal{S}_{\mathbf{k}})=\mathbb{R}$ for $\mathbf{k}\neq 0$.\\
 We define the Green's function matrices as
\begin{equation}
\begin{split}
G_T(z,n,m) &= \langle (\ri \tau \mathcal{S}_{\mathbf{k}}-\mathbb{B}_{8,r}-z)^{-1}e_n,e_m\rangle_{\mathbf{v}},\\
G_S(z,n,m) &= \langle (\ri\tau\mathcal{S}_{\mathbf{k}}-z)^{-1}e_n,e_m\rangle_{\mathbf{v}},
\end{split}
\end{equation}
for $0\leq n,m \leq 4$ and set $G_S(z)=\{G_S(z,n,m)\}_{0\leq n\leq 4}$, $G_T(z)=\{G_T(z,n,m)\}_{0\leq n\leq 4}$.\\
By the second resolvent identity,
\begin{equation}
\mathcal{R}(z;A)-\mathcal{R}(z;B)=\mathcal{R}(z;A)(B-A)\mathcal{R}(z;B),
\end{equation}
for any operators $A,B$ and $z\in\rho(A)\cap\rho(B)$, we have for $A=\ri\tau\mathcal{S}_{\mathbf{k}}$ and $B=\ri\tau\mathcal{S}_{\mathbf{k}}-\mathbb{B}_{8,r}$ that
\begin{equation}\label{res1}
(\ri\tau\mathcal{S}_{\mathbf{k}}-\mathbb{B}_{8,r}-z)^{-1}=(\ri\tau\mathcal{S}_{\mathbf{k}}-z)^{-1}+(\ri\tau\mathcal{S}_{\mathbf{k}}-z)^{-1}\mathbb{B}_{8,r}(\ri\tau\mathcal{S}_{\mathbf{k}}-\mathbb{B}_{8,r}-z)^{-1}.
\end{equation}
Applying equation \eqref{res1} to $e_m$ for $0\leq m\leq 4$ and rearranging gives
\begin{equation}\label{resdelta}
\begin{split}
(\ri\tau\mathcal{S}_{\mathbf{k}}-\mathbb{B}_{8,r}-z)^{-1}e_n&=(\ri\tau\mathcal{S}_{\mathbf{k}}-z)^{-1}e_n+ (\ri\tau\mathcal{S}_{\mathbf{k}}-z)^{-1}\mathbb{B}_{8,r}(\ri\tau\mathcal{S}_{\mathbf{k}}-\mathbb{B}_{8,r}-z)^{-1}e_n\\
&=(\ri\tau\mathcal{S}_{\mathbf{k}}-z)^{-1}e_n+(\ri\tau\mathcal{S}_{\mathbf{k}}-z)^{-1}\sum_{j=0}^7\langle(\ri\tau\mathcal{S}_{\mathbf{k}}-\mathbb{B}_{8,r}-z)^{-1}e_n,e_j\rangle_{\mathbf{v}}(\mathbf{D}_r\mathbf{e})_j\\
&=(\ri\tau\mathcal{S}_{\mathbf{k}}-z)^{-1}e_n+\sum_{j=0}^7G_T(z,n,j)(\ri\tau\mathcal{S}_{\mathbf{k}}-z)^{-1}(\mathbf{D}_r\mathbf{e})_j,
\end{split}
\end{equation} 
for $z\in\mathbb{C}\setminus\ri\mathbb{R}$. Thus, the resolvent of $(\ri\tau\mathcal{S}_{\mathbf{k}}-\mathbb{B}_{8,r}-z)$ includes the resolvent of $\ri\tau\mathcal{S}_{\mathbf{k}}$ as well as information from the matrix $\{G_T(z,n,m)\}_{0\leq n,m\leq 7}$ as coefficients.\\
Taking an inner product of \eqref{resdelta} with $e_m$ gives
\begin{equation}\label{eqGSGT}
\begin{split}
G_T(z,n,m)&=G_S(z,n,m)+\sum_{j=0}^7 G_T(z,n,j)\langle (\ri\tau\mathcal{S}_{\mathbf{k}}-z)^{-1}(\mathbf{D}_r\mathbf{e})_j,e_m \rangle_{\mathbf{v}}\\
&=G_S(z,n,m)+\sum_{j=0}^{7}G_T(z,n,j)\mathbf{D}_rG_S(z,j,m)
\end{split}
\end{equation}
for $0\leq n,m\leq 7$ and $z\in\mathbb{C}\setminus\ri\mathbb{R}$. System \eqref{eqGSGT} defines sixty-four equations for the coefficients $G_T(z,n,m)$, which can be re-written more compactly as
\begin{equation}
G_T=G_S+G_T\mathbf{D}_rG_S,
\end{equation}
or, equivalently,
\begin{equation}\label{IdG_S}
G_T(\Id-\mathbf{D}_rG_S)=G_S.
\end{equation}
Consequently, we can solve for the entries of $G_T$ unless $\det(\Id-\mathbf{D}_rG_S)=0$, which implies that $\sigma_{\rm disc}(\ri \tau\mathcal{S}_k-\mathbb{B}_{8,r})=\{z\in\mathbb{C}: \det(\Id-\mathbf{D}_rG_S(z))=0\}$, which implies the explicit formula
\begin{equation}\label{spec1}
\sigma_{\rm disc}(\hat{\mathcal{L}}_{\mathbf{k}})=-\frac{1}{\tau}-\frac{1}{\tau}\left\{z\in\mathbb{C}:\det\left(\int_{\mathbb{R}^3}\mathbf{D}_r\mathbf{e}(\mathbf{v})\otimes \mathbf{e}(\mathbf{v}) \frac{e^{-\frac{|\mathbf{v}|^2}{2}}}{\ri\tau \mathbf{k}\cdot\mathbf{v}-z}\, \frac{d\mathbf{v}}{(2\pi)^{\frac{3}{2}}}-\Id\right)=0\right\}.
\end{equation}
An eigenvalue $\lambda$ of the operator $\hat{\mathcal{L}}_{\mathbf{k}}$ is then related to the zero $z$ in \eqref{spec1} via
\begin{equation}\label{lambdaz}
z=-\tau\lambda-1.
\end{equation}
To ease notation, we define the \textit{spectral function}
\begin{equation}\label{defSigma}
    \Sigma_{\mathbf{k},\tau}(\lambda) = \det\left(\int_{\mathbb{R}^3}\mathbf{D}_r\mathbf{e}(\mathbf{v})\otimes \mathbf{e}(\mathbf{v}) \frac{e^{-\frac{|\mathbf{v}|^2}{2}}}{\ri\tau \mathbf{k}\cdot\mathbf{v}+\lambda \tau +1}\, \frac{d\mathbf{v}}{(2\pi)^{\frac{3}{2}}}-\Id\right),
\end{equation}
such that 
\begin{equation}
    \sigma_{\rm disc}(\hat{\mathcal{L}}_{\mathbf{k}}) = \{\lambda \in \mathbb{C}: \Sigma_{\mathbf{k},\tau}(\lambda) = 0\}. 
\end{equation}

\subsection{Derivation of a Spectral Function}\label{spectral2}
To evaluate the integral expression in \eqref{defSigma}, we we decompose the wave vector $\mathbf{k}$ into its magnitude along a coordinate direction and a rotation:
\begin{equation}
\mathbf{k}=\mathbf{Q}_{\mathbf{k}}(k,0,0)^T,
\end{equation}
where $Q_{\mathbf{k}}Q_{\mathbf{k}}^T=\Id$. Setting $\mathbf{w}=\mathbf{Q}_{\mathbf{k}}^T\mathbf{v}$, we have that
\begin{equation}\label{vw}
\mathbf{k}\cdot\mathbf{v}=\mathbf{Q}_{\mathbf{k}}(|\mathbf{k}|,0,0)^T\cdot\mathbf{v} = (k,0,0)\cdot\mathbf{w} = kw_1,
\end{equation}
while the vector of basis functions $\mathbf{e}$ transforms according to 
\begin{equation}
\begin{split}
\mathbf{e}(\mathbf{v})&=\left(1,\mathbf{v},\frac{|\mathbf{v}|^2-3}{\sqrt{6}},\mathbf{v}\frac{|\mathbf{v}|^2-5}{\sqrt{10}}\right)=\left(1,\mathbf{Q}_{\mathbf{k}}\mathbf{w},\frac{|\mathbf{w}|^2-3}{\sqrt{6}},\mathbf{Q}_{\mathbf{k}}\mathbf{w}\frac{|\mathbf{w}|^2-5}{\sqrt{10}}\right)\\
&=
\begin{pmatrix}
1 & 0 & 0 & 0\\
0 & \mathbf{Q}_{\mathbf{k}} & 0 & 0\\
0 & 0 & 1 & 0\\
0 & 0 & 0 & \mathbf{Q}_{\mathbf{k}}
\end{pmatrix}\mathbf{e}(\mathbf{w}).
\end{split}
\end{equation}
\begin{remark}
   The first column of the rotation matrix $\mathbf{Q}_{\mathbf{k}}$ can be chosen as $\frac{1}{k}\mathbf{k}$, which, by the orthonormality of the columns of $\mathbf{Q}_{\mathbf{k}}$, implies \eqref{vw}. The change of coordinates $\mathbf{v}=\mathbf{Q}_{\mathbf{k}}\mathbf{w}$ can then interpreted as writing the velocity vector $\mathbf{v}$ as the sum of a a component parallel to the wave vector $\mathbf{k}$ and a component orthogonal to it: $\mathbf{v}= \mathbf{v}_{\parallel}+\mathbf{v}_{\perp}$ with $\mathbf{v}_{\parallel}=\frac{\mathbf{v}\cdot\mathbf{k}}{k^2}\mathbf{k}$. 
\end{remark}

By the orthogonality of $\mathbf{Q}_{\mathbf{k}}$, the volume element transforms as  $d\mathbf{v}=d\mathbf{w}$ and we can calculate
\begin{equation}\label{det1}
\begin{split}
&\det\left(\int_{\mathbb{R}^3}\mathbf{D}_r\mathbf{e}(\mathbf{v})\otimes\mathbf{e}(\mathbf{v}) \frac{e^{-\frac{|\mathbf{v}|^2}{2}}}{\ri\tau \mathbf{k}\cdot\mathbf{v}-z}\, \frac{d\mathbf{v}}{(2\pi)^{\frac{3}{2}}}-\Id\right)\\
&\qquad=\det\left(\int_{\mathbb{R}^3}\left(\mathbf{D}_r\begin{pmatrix}
1 & 0 & 0 & 0\\
0 & \mathbf{Q}_{\mathbf{k}} & 0 & 0\\
0 & 0 & 1 & 0\\
0 & 0 & 0 & \mathbf{Q}_{\mathbf{k}}
\end{pmatrix}\mathbf{e}(\mathbf{w})\right)\otimes \begin{pmatrix}
1 & 0 & 0 & 0\\
0 & \mathbf{Q}_{\mathbf{k}} & 0 & 0\\
0 & 0 & 1 & 0\\
0 & 0 & 0 & \mathbf{Q}_{\mathbf{k}}
\end{pmatrix}\mathbf{e}(\mathbf{w})\frac{e^{-\frac{|\mathbf{w}|^2}{2}}}{\ri\tau kw_1-z}\, \frac{d\mathbf{w}}{(2\pi)^{\frac{3}{2}}}-\Id\right)\\
&\qquad=\det\left(\int_{\mathbb{R}^3}\begin{pmatrix}
1 & 0 & 0 & 0\\
0 & \mathbf{Q}_{\mathbf{k}} & 0 & 0\\
0 & 0 & 1 & 0\\
0 & 0 & 0 & \mathbf{Q}_{\mathbf{k}}
\end{pmatrix}\mathbf{D}_r\mathbf{e}(\mathbf{w})\otimes \left(\begin{pmatrix}
1 & 0 & 0 & 0\\
0 & \mathbf{Q}_{\mathbf{k}} & 0 & 0\\
0 & 0 & 1 & 0\\
0 & 0 & 0 & \mathbf{Q}_{\mathbf{k}}
\end{pmatrix}\mathbf{e}(\mathbf{w})\right)\frac{e^{-\frac{|\mathbf{w}|^2}{2}}}{\ri\tau kw_1-z}\, \frac{d\mathbf{w}}{(2\pi)^{\frac{3}{2}}}-\Id\right)\\
&\qquad=\det\left(\int_{\mathbb{R}^3}\mathbf{D}_r\mathbf{e}(\mathbf{w})\otimes\mathbf{e}(\mathbf{w}) \frac{e^{-\frac{|\mathbf{w}|^2}{2}}}{\ri\tau kw_1-z}\, \frac{d\mathbf{w}}{(2\pi)^{\frac{3}{2}}}-\Id\right),\\
\end{split}
\end{equation}
where we have used that $\mathbf{D}_r$ only acts on the last three columns by multiplication with a constant and hence commutes with the block-rotation-matrix, together with the orthogonality of $\mathbf{Q}_{\mathbf{k}}$ in factoring the determinant.\\
From equation \eqref{det1}, we see that the spectral function $\Sigma_{\mathbf{k},\tau}$ depends on the wave vector $\mathbf{k}$ only through $\tau k$ and we define
\begin{equation}
    \kappa := \tau k. 
\end{equation}
A lengthy but elementary calculations allows us to integrate out the variables $w_2$ and $w_3$ in \eqref{det1}, which simplifies the spectral function according to
\begin{equation}\label{det2}
\begin{split}
\det\left(\int_{\mathbb{R}^3}\mathbf{D}_r\mathbf{e}(\mathbf{v})\otimes\mathbf{e}(\mathbf{v}) \frac{e^{-\frac{|\mathbf{v}|^2}{2}}}{\ri\tau \mathbf{k}\cdot\mathbf{v}-z}\, \frac{d\mathbf{v}}{(2\pi)^{\frac{3}{2}}}-\Id\right) &=  \det\left(\int_{\mathbb{R}^3}\mathbf{D}_r\mathbf{M}(w)\frac{e^{-\frac{w^2}{2}}}{\ri\kappa w-z}\, \frac{dw}{\sqrt{2\pi}}-\Id\right)\\
& = \frac{1}{(\ri\kappa )^8}\det\left(\int_{\mathbb{R}^3}\mathbf{D}_r\mathbf{M}(w)\frac{e^{-\frac{w^2}{2}}}{w-\zeta}\, \frac{dw}{\sqrt{2\pi}}-(\ri\kappa )\Id\right)_{\zeta=\frac{z}{\ri\kappa }},
\end{split}
\end{equation}
where
\begin{equation}\label{defM}
    \mathbf{M}(w) = \begin{pmatrix}
        \mathbf{M}_{11}(w) & \mathbf{M}_{12}(w) \\
        \mathbf{M}_{12}^T(w) & \mathbf{M}_{22}(w)
    \end{pmatrix},
\end{equation}
and the coefficient matrices are given by
\begin{equation}\label{defM2}
    \begin{split}
       \mathbf{M}_{11}(w) & = \begin{pmatrix}
           1 & 0 & 0 & w\\
           0 & 1 & 0 & 0  \\
           0 & 0 & 1 & 0\\
           w & 0 & 0 & w^2
       \end{pmatrix},\quad  \mathbf{M}_{12}(w) = \begin{pmatrix} \frac{w^2-1}{\sqrt{6}} & 0 & 0 & \frac{ w \left(w^2-3\right)}{\sqrt{10}}\\
       0 & \frac{ \left(w^2-1\right)}{\sqrt{10}} & 0 & 0\\
       0 & 0 & \frac{ \left(w^2-1\right)}{\sqrt{10}} & 0 \\
       \frac{w \left(w^2-1\right)}{\sqrt{6}} & 0 & 0 & \frac{ w^2
   \left(w^2-3\right)}{\sqrt{10}} \end{pmatrix},\\
   \mathbf{M}_{22}(w) & = \begin{pmatrix}
 \frac{1}{6} \left(w^4-2
   w^2+5\right) & 0 & 0 & \frac{ w \left(w^4-4 w^2+7\right)}{2 \sqrt{15}} \\
  0 & \frac{1}{10}  \left(w^4-2 w^2+9\right) & 0 & 0 \\
  0 & 0 & \frac{1}{10}  \left(w^4-2 w^2+9\right) & 0 \\
   \frac{w \left(w^4-4 w^2+7\right)}{2 \sqrt{15}} & 0 & 0 & \frac{1}{10}  w^2 \left(w^4-6
   w^2+13\right)
\end{pmatrix}
    \end{split}
\end{equation}
Since the entries of $\mathbf{M}$ are polynomials of degree at most six, we can expand
\begin{equation}
    \mathbf{M}(w) = \sum_{n=0}^6 \mathbf{M}_{n} w^n,
\end{equation}
for some matrix coefficients $\mathbf{M}_{n}\in\mathbb{R}^{8\times 8}$.\\ 
To evaluate the integral expression on the right-hand side of \eqref{det2}, we will use the plasma dispersion function \eqref{defZ}. To this end, we have to calculate expressions for the derivative of $Z$. By repeated application of \eqref{diffZ}, we conclude that 
\begin{equation}
    \frac{d^nZ}{dz^n}(\zeta) = p_n(\zeta) + q_{n}(\zeta)Z(\zeta), \qquad n\geq 1 
\end{equation}
for some polynomials $p_n,q_n$ with integer coefficients. The first few derivatives of $Z$ can be expressed as
\begin{equation}\label{diffI0}
\begin{split}
\frac{dZ}{d\zeta}&=-1-\zeta Z,\\
\frac{d^2Z}{d\zeta^2}&=\zeta+(\zeta^2-1)Z,\\
\frac{d^3Z}{d\zeta^3}&=2-\zeta^2+(3\zeta-\zeta^2)Z,\\
\frac{d^4Z}{d\zeta^4}&=-5\zeta+\zeta^3+(\zeta^4-6\zeta^2+3)Z.
\end{split}
\end{equation}
We claim that $q_n(\zeta) = (-1)^nH_n(\zeta)$ for the $n^{th}$ Hermite polynomial. Indeed, from the recurrence relation of $Z$ in \eqref{diffZ}, we have that 
\begin{equation}
\begin{split}
    p_{n+1} + q_{n+1} Z &= \frac{d^{n+1}Z}{d\zeta^{n+1}} = \frac{d}{d\zeta}(p_n+q_nZ) = p'_n+q'_nZ+q_nZ'\\
    & = p'_n+q'_nZ+q_n(-\zeta Z-1) = (p'_n-q_n) + (q'_n-\zeta q_n)Z,
    \end{split}
\end{equation}
showing that $q_{n+1} = q'_n-\zeta q_n$, which is - up to a sign flip - the recurrence relation of the $n^{th}$ Hermite polynomial \eqref{recHermite2}.  With the relation
\begin{equation}\label{HermiteInt}
\begin{split}
\frac{1}{\sqrt{2\pi}}\int_{\mathbb{R}} H_k(v)\frac{e^{-\frac{v^2}{2}}}{v-\zeta}\, dv& = \frac{1}{\sqrt{2\pi}}\int_{\mathbb{R}}\left[\left(-\frac{d}{dv}\right)^ke^{-\frac{v^2}{2}}\right]\frac{dv}{v-\zeta}= \frac{(-1)^kk!}{\sqrt{2\pi}}\int_{\mathbb{R}}e^{-\frac{v^2}{2}}\frac{dv}{(v-\zeta)^{k+1}}\\
&=\frac{(-1)^k}{\sqrt{2\pi}}\frac{d^k}{d\zeta^k}\int_{\mathbb{R}}e^{-\frac{v^2}{2}}\frac{dv}{v-\zeta}= (-1)^k\frac{d^kZ}{d\zeta^k},
\end{split}
\end{equation}
we can now conclude that for any polynomial expanded in Hermite basis $P(w) =\sum_{n=0}^N P_n H_n(w)$:
\begin{equation}\label{insight}
\begin{split}
\int_{\mathbb{R}} P(w) \frac{e^{-\frac{w^2}{2}}}{w-\zeta}\, dw & = \sum_{n=0}^NP_n\int_{\mathbb{R}} H_n(w)\frac{e^{-\frac{w^2}{2}}}{w-\zeta}\, dw = \sum_{n=0}^NP_n(-1)^n\frac{d^nZ}{d\zeta^n}(\zeta)\\
& = \sum_{n=0}^NP_n(-1)^n(p_n(\zeta) + (-1)^nH_n(\zeta) Z(\zeta))\\
& = P(\zeta) Z(\zeta) + P\sum_{n=0}^NP_n(-1)^np_n(\zeta) .
\end{split}
\end{equation}
With the help of the plasma dispersion function $Z$ and the insight \eqref{insight}, we readily calculate 
\begin{equation}\label{expansionZ}
\begin{split}
\int_{\mathbb{R}} \mathbf{M}(w)\frac{e^{-\frac{w^2}{2}}}{w-\zeta}\, dw - \mathbf{M}(\zeta) Z(\zeta) & = \int_{\mathbb{R}} [\mathbf{M}(w)-\mathbf{M}(\zeta)] \frac{e^{-\frac{w^2}{2}}}{w-\zeta}\, dw\\
& = \sum_{n=0}^6 \mathbf{M}_{n}\int_{\mathbb{R}} \frac{w^n-\zeta^n}{w-\zeta}e^{-\frac{w^2}{2}}\, dw\\
& = \sum_{n=0}^6\sum_{j=0}^{n-1}\mathbf{M}_{n}\left(\int_{\mathbb{R}} w^{n-j-1}e^{-\frac{w^2}{2}}\, dw \right)\zeta^j.
\end{split}
\end{equation}
Using \eqref{expansionZ}, we can evaluate the integral expression in \eqref{det1} to an explicit matrix depending on polynomials in $\zeta$ and $Z(\zeta)$ in a linear way. Indeed, with the help of \eqref{expansionZ}, we have that
\begin{equation}
    \begin{split}
        \det\left(\int_{\mathbb{R}^3}\mathbf{D}_r\mathbf{M}(w)\frac{e^{-\frac{w^2}{2}}}{w-\zeta}\, \frac{dw}{\sqrt{2\pi}}-(\ri\kappa )\Id\right)_{\zeta=\frac{z}{\ri\kappa }} = \det\left(\mathbf{D}_r\mathbf{N}(\zeta) -(\ri\kappa )\Id\right),
    \end{split}
\end{equation}
for the matrix
\begin{equation}
    \mathbf{N}(\zeta) = 
\begin{pmatrix}
        \mathbf{N}_{11}(\zeta) & \mathbf{N}_{12}(\zeta) \\
        \mathbf{N}_{12}^T(\zeta) & \mathbf{N}_{22}(\zeta)
    \end{pmatrix},
\end{equation}
and the coefficient matrices are given by
\begin{equation}\label{Zzetaexplicit}
    \begin{split}
       \mathbf{N}_{11}(\zeta) & = \begin{pmatrix}
            Z(\zeta )   & 0 & 0 & \zeta  Z(\zeta )+1  \\
            0 & Z(\zeta )  & 0 & 0  \\
           0 & 0 & Z(\zeta )   & 0\\
            \zeta  Z(\zeta )+1 & 0 & 0 & \zeta  +\zeta ^2 Z(\zeta )
       \end{pmatrix},\\
       \mathbf{N}_{12}(\zeta)  & = \begin{pmatrix} 
       \frac{\zeta }{\sqrt{6}}+\frac{\left(\zeta ^2-1\right) Z(\zeta )}{\sqrt{6}} & 0 & 0 &
   \frac{\left(\zeta ^2-2\right) }{\sqrt{10}}+\frac{\zeta  \left(\zeta ^2-3\right)  Z(\zeta )}{\sqrt{10}}\\
       0 & \frac{\zeta  }{\sqrt{10}}+\frac{\left(\zeta ^2-1\right)  Z(\zeta )}{\sqrt{10}} & 0 & 0 \\
       0 & 0 & \frac{\zeta  }{\sqrt{10}}+\frac{\left(\zeta ^2-1\right)  Z(\zeta )}{\sqrt{10}} & 0 \\
       \frac{\zeta ^2}{\sqrt{6}}+\frac{\left(\zeta ^2-1\right) \zeta 
   Z(\zeta )}{\sqrt{6}} & 0 & 0 & \frac{\left(\zeta ^2-2\right) \zeta  }{\sqrt{10}}+\frac{\left(\zeta ^2-3\right) \zeta ^2  Z(\zeta
   )}{\sqrt{10}}
        \end{pmatrix},\\
   \mathbf{N}_{22}(\zeta) & = \begin{pmatrix}
  N_{22,1} & 0 & 0 & \frac{\left(\zeta ^4-3 \zeta ^2+6\right) }{2 \sqrt{15}}+\frac{\zeta  \left(\zeta ^4-4 \zeta
   ^2+7\right)  Z(\zeta )}{2 \sqrt{15}}\\
 0 &  N_{22,2}  & 0 &  0  \\
 0 & 0 &  N_{22,3}  & 0 \\
   \frac{\zeta ^4-3 \zeta ^2+6}{2 \sqrt{15}}+\frac{\zeta 
   \left(\zeta ^4-4 \zeta ^2+7\right) Z(\zeta )}{2 \sqrt{15}} & 0 & 0 &  N_{22,4}
\end{pmatrix},\\
N_{22,1} & = \frac{1}{6} \zeta  \left(\zeta ^2-1\right)  +\frac{1}{6} \left(\zeta ^4-2 \zeta
   ^2+5\right) Z(\zeta )\\
N_{22,2} & = \frac{1}{10} \zeta 
   \left(\zeta ^2-1\right) +\frac{1}{10} \left(\zeta ^4-2 \zeta ^2+9\right)  Z(\zeta )\\
N_{22,3} & =\frac{1}{10} \zeta 
   \left(\zeta ^2-1\right) +\frac{1}{10} \left(\zeta ^4-2 \zeta ^2+9\right)  Z(\zeta )\\
N_{22,4} & = \frac{1}{10} \left(\zeta ^4-5 \zeta ^2+10\right)
   \zeta  +\frac{1}{10} \left(\zeta ^4-6 \zeta ^2+13\right) \zeta ^2  Z(\zeta ).
    \end{split}
\end{equation}
Using the change of coordinates 
\begin{equation}
\zeta=\frac{-\tau \lambda-1}{\ri\kappa } = \ri\frac{\tau \lambda+1}{\kappa},
\end{equation}
the spectral function $\Sigma_{\mathbf{k},\tau}$ takes the explicit form
\begin{equation}\label{Sigmaexpl}
\begin{split}
    \Sigma_{\mathbf{k},\tau}(\lambda) & = \frac{1}{3000(\ri \kappa )^8}\Big[ \Sigma_0(\zeta) +\Sigma_1(\zeta)Z(\zeta )+ \Sigma_2(\zeta) Z(\zeta )^2\Big]^2 \Big[\Sigma_3(\zeta)+\Sigma_4(\zeta)Z(\zeta)+\Sigma_5(\zeta)Z^2(\zeta)\Big]_{\zeta=\ri\frac{\tau \lambda+1}{\kappa}},
   \end{split}
\end{equation}
for the polynomials
\begin{equation}\label{coefpoly}
    \begin{split}
        \Sigma_0(\zeta) & = 10 \kappa ^2+\zeta  r \left(\ri \zeta ^2 \kappa +\zeta -\ri \kappa \right),\\
        \Sigma_1(\zeta) & = 10 \ri \kappa +r \left(\ri \zeta ^4 \kappa +\zeta ^3-2 \ri \zeta ^2
   \kappa -\zeta +9 \ri \kappa \right),\\
   \Sigma_2(\zeta) &= -8r,\\
   \Sigma_3(\zeta)& = 5 \kappa  \left(\ri \zeta ^3 \kappa ^2+2 \zeta ^2 \kappa +\ri \zeta  \left(5 \kappa ^2-1\right)+6 \left(\kappa ^3+\kappa \right)\right)\\
   &\qquad+r \left(3
   \ri \zeta ^5 \kappa ^3+9 \zeta ^4 \kappa ^2-3 \ri \zeta ^3 \kappa  \left(5 \kappa ^2+3\right)-\zeta ^2 \left(43 \kappa ^2+3\right)+2 \ri \zeta 
   \kappa  \left(15 \kappa ^2+23\right)+6 \left(5 \kappa ^2+3\right)\right),\\
   \Sigma_4(\zeta)  &= \ri \left(5 \kappa  \left(\zeta ^4 \kappa ^2-2 \ri \zeta ^3 \kappa +\zeta ^2 \left(4 \kappa ^2-1\right)+11 \kappa ^2+5\right)+r \left(3 \zeta ^6
   \kappa ^3-9 \ri \zeta ^5 \kappa ^2-9 \zeta ^4 \left(2 \kappa ^3+\kappa \right)\right.\right.\\
   &\qquad+\left.\left.\ri \zeta ^3 \left(64 \kappa ^2+3\right)+\zeta ^2 \kappa 
   \left(39 \kappa ^2+79\right)-\ri \zeta  \left(23 \kappa ^2+33\right)+16 \kappa \right)\right),\\
   \Sigma_5(\zeta) & = -4 \left(5 \kappa  \left(\zeta ^2 \kappa -\ri \zeta +\kappa \right)+\zeta  r \left(3 \zeta ^3 \kappa ^2-6 i \zeta ^2 \kappa +5 \zeta  \kappa
   ^2-3 \zeta -5 \ri \kappa \right)\right).
    \end{split}
\end{equation}
Equation \eqref{Sigmaexpl} shows that the spectral function factors into two parts, which we denote as
\begin{equation}\label{factorization}
    \begin{split}
        \Sigma_{\rm shear}(\zeta) := \frac{1}{10\kappa^2}[\Sigma_0(\zeta) +\Sigma_1(\zeta)Z(\zeta )+ \Sigma_2(\zeta) Z^2(\zeta )],\\
        \Sigma_{\rm diff,ac}(\zeta) := \frac{1}{30\kappa^4}[\Sigma_3(\zeta)+\Sigma_4(\zeta)Z(\zeta)+\Sigma_5(\zeta)Z^2(\zeta)].
    \end{split}
\end{equation}
Let us conclude this section with some remarks. The main result of the preceding calculations - the derivation of the spectral function \eqref{Sigmaexpl} with coefficient polynomials and factorization \eqref{factorization} - allows us to conclude specific features of the spectrum by solving for the zeros of a holomorphic function. This is a tremendous simplification compared to the study of \eqref{defL} directly, where the transport and the collision term interact in a delicate manner. This is as close an analog of a determinant in finite-dimensional systems as one could hope for.\\
In the following section, we will take a closer look at the structure of the zero set of \eqref{Sigmaexpl}. In particular, we will identify different families of zeros (branches) that relate to hydrodynamics. 


\subsection{Hydrodynamic Modes}\label{spectral3}
\begin{figure}
     \centering
     \begin{subfigure}[b]{0.45\textwidth}
         \centering
         \includegraphics[width=\textwidth]{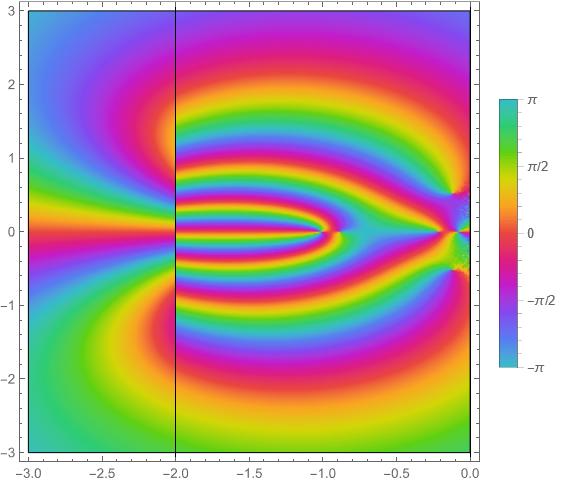}
         \caption{k= 0.4}
     \end{subfigure}
     \begin{subfigure}[b]{0.45\textwidth}
         \centering
         \includegraphics[width=\textwidth]{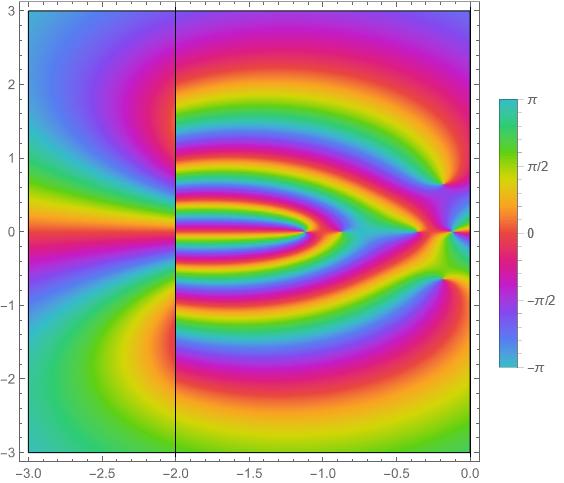}
         \caption{k=0.5}
     \end{subfigure}
     \begin{subfigure}[b]{0.45\textwidth}
         \centering
         \includegraphics[width=\textwidth]{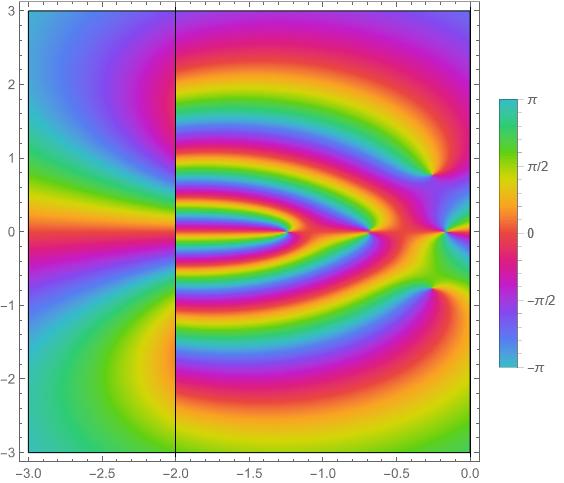}
         \caption{k=0.6}
     \end{subfigure}
     \begin{subfigure}[b]{0.45\textwidth}
         \centering
         \includegraphics[width=\textwidth]{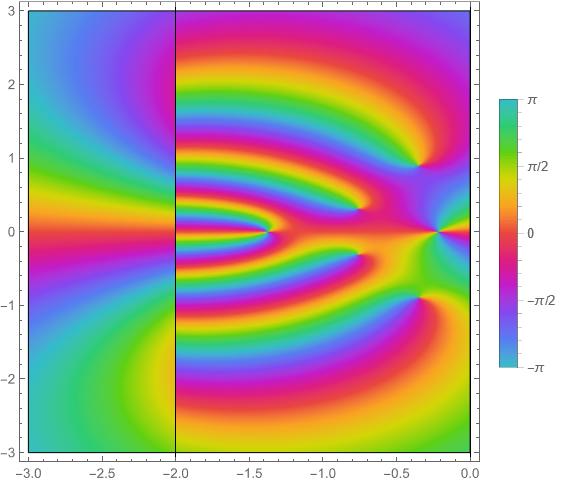}
         \caption{k=0.7}
     \end{subfigure}
        \caption{Argument plot of the spectral function \eqref{defSigma} for relaxation time $\tau = 0.5$, Prandtl number $Pr=0.4$ and different wave numbers $k$. The zeros of \eqref{defSigma} define eigenvalues of the linearized Shakhov mode (points where a small, counter-clockwise loop runs through the whole rainbow at least once). All eigenvalues have negative real part and are located above the essential spectrum $\{\Re\lambda = -\frac{1}{\tau}\}$ (solid black line), which is consistent with the decay estimates \eqref{lowerbound}.\\
        At small wave numbers ($k=0.4$ and $k=0.5$), the primary diffusion mode decreases along the real axis, while the secondary diffusion mode increases along the real axis. Around $k\approx 0.6$ they collide, a bifurcation takes place and a pair of complex conjugated modes, the secondary acoustics, is created ($k=0.7$).\\
        \medskip}
        \label{specplot22}
\end{figure}

In this section, we will identify the branches of the zero set of the spectral function $\Sigma_{\mathbf{k},\tau}$ in dependence on the modified wave number $\kappa$ and the parameter $r$.\\
First, let us show that in the limit $\kappa\to 0$ (for fixed $\tau$), we recover the spectral structure of \eqref{sigmaL0}. To this end, we use the asymptotic expansion
\begin{equation}\label{Zasymptotic}
    Z(\zeta) \sim -\sum_{n=0}^\infty \frac{(2n-1)!!}{\zeta^{2n+1}},\quad \text{ for } |\arg(\zeta)|\leq \frac{\pi}{2}-\delta,\quad \zeta\to\infty,
\end{equation}
under the assumption that $\Im\zeta>0$. For a proof of \eqref{Zasymptotic}, we refer to the Appendix. The limit $\kappa\to 0$ with $\Re\lambda < 0$ satisfies the assumptions of \eqref{Zasymptotic}. A lengthy calculation shows that, indeed, 
\begin{equation}
    \begin{split}
        \Sigma_{\mathbf{k},\tau}(\lambda) \sim \frac{\lambda^5 \tau ^{5} (\lambda \tau +1-r)^3}{(\lambda \tau +1)^8}, \quad \text{ for } \kappa\to 0 \text{ and } \Re\lambda>0,
    \end{split}
\end{equation}
which is consistent with \eqref{sigmaL0}. Since, for small $\kappa$, the spectral function $\Sigma_{\mathbf{k},\tau}$ is continuous in $\kappa$, it follows that there exist five branches of eigenvalues (indexed by $\kappa$) emerging out of the five-fold degenerate zero $\lambda = 0$, which we call \textit{primary hydrodynamic modes}. Furthermore, there exist three branches of eigenvalues emerging from the zero $\lambda = -\frac{1}{\tau_{\rm slow}}$ which we call \textit{secondary hydrodynamic modes}. From the factorization in \eqref{Sigmaexpl} and \eqref{factorization}, we see that any zero of $\Sigma_1$ will occur with algebraic multiplicity of two.\\
For small wave number, the primary hydrodynamic modes consist of a pair of complex conjugate eigenvalues, the \textit{primary acoustic modes}, a real eigenvalue of algebraic multiplicity two (but geometric multiplicity one), called \textit{primary shear mode} and a algebraically and geometrically simple, real eigenvalue, called \textit{primary diffusion mode}. Similarly, for small wave number, the secondary hydrodynamic modes consist of a real eigenvalue of algebraic multiplicity two (but geometric multiplicity one), called \textit{secondary shear mode} and a algebraically and geometrically simple, real eigenvalue, called \textit{secondary diffusion mode}, see Figure \ref{specplot22}.\\
For larger wave numbers, depending on $Pr$, the primary and secondary shear modes may collide and produce another pair of complex conjugated eigenvalues, called \textit{secondary acoustic modes} or \textit{second sound}, see Figure \ref{specplot22}. This occurs through a saddle-node bifurcation, see Section \ref{spectral4}.\\
We denote the families of modes index by wave number $k$ as
 \begin{equation}
     \kappa \mapsto \lambda_{N}(\kappa),\quad \text{ for } N\in {\rm Modes}(\kappa),
 \end{equation}
 where the low wave-number set of modes is given by
 \begin{equation}\label{mode1}
    \rm Modes_1 = \{\rm shear_1,diff_1,ac_1,ac_1*,shear_2,diff_2\},
 \end{equation}
 while the higher wave-number set of modes is given by
  \begin{equation}\label{mode2}
     \rm Modes_2 = \{\rm shear_1,ac_1,ac_1*,shear_2,ac_2,ac_2*\}.
 \end{equation}
 As mentioned before, in a certain range of Prandtl numbers, the set of modes can change from \eqref{mode1} to \eqref{mode2} if the wave number is increased. 

\subsection{Existence of a Critical Wave Number and Finiteness of the Hydrodynamic Spectrum}\label{spectral3b}

Along the same lines as in \cite{kogelbauer2023exact}, we will show that for each family of modes, there exists a critical wave number such that $k_{\rm crit}$ such that 
\begin{equation}
    \sigma_{\rm disc}(\hat{\mathcal{L}}_{\mathbf{k}})=\emptyset,\quad \text{ for } |\mathbf{k}|> k_{\rm crit}. 
\end{equation}
In fact, each mode has its own critical wave number, which depends on the specific properties of the branch.s\\
\begin{proof}
The claim follows from a combination of Rouch{\'e}'s theorem applied to the spectral function $\Sigma_{\mathbf{k},\tau}$ with the asymptotic expansion \eqref{Zasymptotic}. Indeed, for fixed $\kappa$, we find that 
\begin{equation}\label{asy1}
\begin{split}
    \Sigma_{shear}(\zeta) & \sim \frac{1}{10\kappa^2}\left[\Sigma_0(\zeta) +\Sigma_1(\zeta)\left( -\frac{1}{\zeta}-\frac{1}{\zeta^3}+\mathcal{O}(\zeta^{-5}) \right) + \Sigma_2(\zeta)\left( -\frac{1}{\zeta}-\frac{1}{\zeta^3}+\mathcal{O}(\zeta^{-5}) \right)\right]\\
    & \sim \frac{10 \zeta ^3 \kappa  \left(\zeta ^3 \kappa -\ri \zeta ^2-i\right)+r \left(-7 \ri \zeta ^5 \kappa -7 \zeta ^4-9 \ri \zeta ^3 \kappa -16 \zeta
   ^2-8\right)}{10\kappa^2\zeta ^6}\\
   &\sim 1,
\end{split}
\end{equation}
$|\arg(\zeta)|\leq \frac{\pi}{2}-\delta, \quad  \zeta\to\infty$, for any real number $0<\delta\leq \frac{\pi}{2}$.\\
Analogously, we find that
\begin{equation}\label{asy2}
\begin{split}
    \Sigma_{diff,ac}(\zeta) & \sim \frac{1}{30\kappa^4}\left[\Sigma_3(\zeta) +\Sigma_4(\zeta)\left( -\frac{1}{\zeta}-\frac{1}{\zeta^3}+\mathcal{O}(\zeta^{-5}) \right) + \Sigma_5(\zeta)\left(-\frac{1}{\zeta}-\frac{1}{\zeta^3}+\mathcal{O}(\zeta^{-5}) \right)\right]\\
    & \sim \frac{\ri}{30\kappa^4\zeta ^{10}}\Big[ 5 \kappa  \left(-6 i \zeta ^{10} \kappa ^3-18 \zeta ^9 \kappa ^2+18 i \zeta ^8 \kappa +\zeta ^7 \left(6-23 \kappa ^2\right)\right.\\
    &\qquad\left.+36 i
   \zeta ^6 \kappa +\zeta ^5 \left(13-33 \kappa ^2\right)+52 i \zeta ^4 \kappa +24 \zeta ^3+60 i \zeta ^2 \kappa +36 \zeta +36 i \kappa
   \right)\\
   &\qquad+\zeta  r \left(15 \zeta ^8 \kappa ^3-45 i \zeta ^7 \kappa ^2-9 \zeta ^6 \kappa  \left(13 \kappa ^2+5\right)+i \zeta ^5 \left(281
   \kappa ^2+15\right)+236 \zeta ^4 \kappa\right.\\
   &\qquad\left.+12 i \zeta ^3 \left(19 \kappa ^2-6\right)+336 \zeta ^2 \kappa +36 i \zeta  \left(5 \kappa
   ^2-3\right)+180 \kappa \right)\Big]\\
   &\sim 1,
\end{split}
\end{equation}
$|\arg(\zeta)|\leq \frac{\pi}{2}-\delta, \quad  \zeta\to\infty$, for any real number $0<\delta\leq \frac{\pi}{2}$.\\
Since $\Sigma_{|\mathbf{k}|,\tau}$ is an analytic function in the strip $\{-\frac{1}{\tau}<\Re\lambda<0\}$ and continuously extends to the boundary, it follows from \eqref{asy1} and \eqref{asy2} that $\lambda\mapsto |\Sigma_{|\mathbf{k}|,\tau}(\lambda)-1|$ is bounded and converges to zero for $\lambda\to \infty$ with $-\frac{1}{\tau}\leq \Re\lambda\leq 0$. Next, we observe that $|\Sigma_{|\mathbf{k}|,\tau}(\lambda)-1|$ only contains terms in $k$ of order $\mathcal{O}(k^{-1})$. This shows that there exists a number $k_{\rm crit}$ such that
\begin{equation}
|\Sigma_{|\mathbf{k}|,\tau}(\lambda)-1|<1,
\end{equation}
for $k>k_{\rm crit}$, uniformly on any rectangle of the form $\mathbf{R}_a=\{- a, a, a+\ri\frac{1}{\tau|\mathbf{k}|},-a+\ri\frac{1}{\tau|\mathbf{k}|} \}$ for $a>0$. Consequently, by Rouch{\'e}'s theorem, since the constant function $1$ does not have any zeros, the spectral function $\lambda\mapsto \Sigma_{|\mathbf{k}|,\tau}(\lambda)$ cannot have any zeros with $-\frac{1}{\tau}\leq \Re\lambda\leq 0$ for $k>k_{\rm crit}$ either. This proves the claim.
\end{proof}

\begin{remark}
The critical wave number obtained before depend inversely on the (non-dimensional) relaxation parameter. Defining the typical length of a mean free path as
\begin{equation}\label{defmeanfree}
    L_{mfp} = \tau v_{thermal},
\end{equation}
and transforming back to physical units, we see that the critical wave number is numerically proportional to the inverse of \eqref{defmeanfree}. Indeed, we obtain that
\begin{equation}
k_{\rm crit}\sim \sqrt{\frac{k_BT_0}{m}}\frac{1}{\tau_{phys}}\sim \frac{1}{l_{mfp}}. 
\end{equation}
\end{remark}

\begin{remark}
In fact, each family of modes has its own critical wave number $k_{crit,N}$. Since branches can merge, 
\end{remark}

\subsection{Global Bifurcation of Eigenvalues, Merging of Branches and Second Sound}\label{spectral4}
\begin{figure}
     \centering
     \begin{subfigure}[b]{0.45\textwidth}
         \centering
         \includegraphics[width=\textwidth]{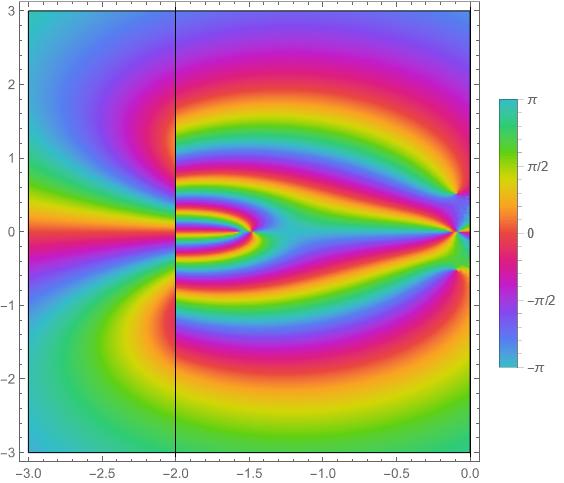}
         \caption{$k=0.4$}
     \end{subfigure}
     \begin{subfigure}[b]{0.45\textwidth}
         \centering
         \includegraphics[width=\textwidth]{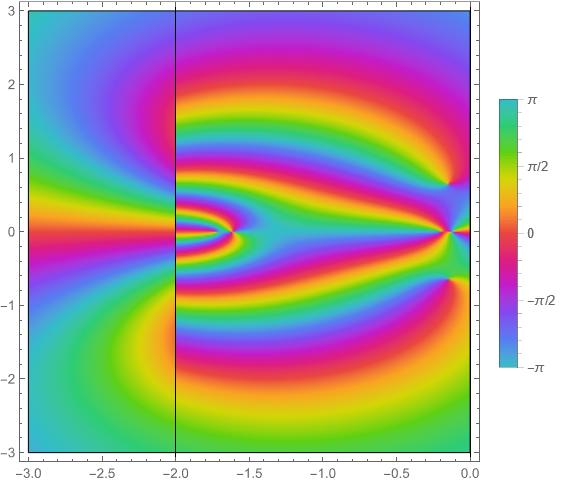}
         \caption{k=0.5}
     \end{subfigure}
     \begin{subfigure}[b]{0.45\textwidth}
         \centering
         \includegraphics[width=\textwidth]{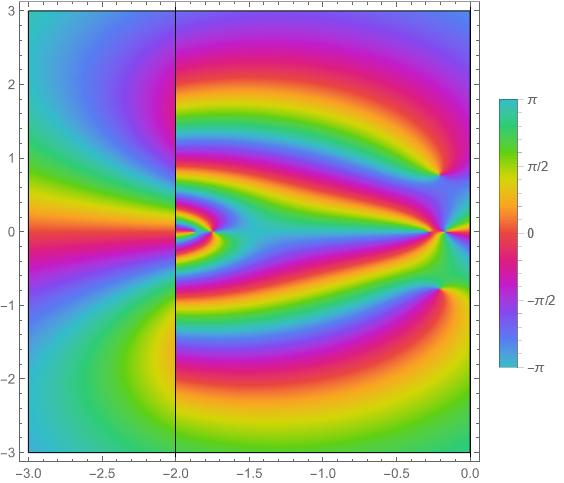}
         \caption{k=0.6}
     \end{subfigure}
     \begin{subfigure}[b]{0.45\textwidth}
         \centering
         \includegraphics[width=\textwidth]{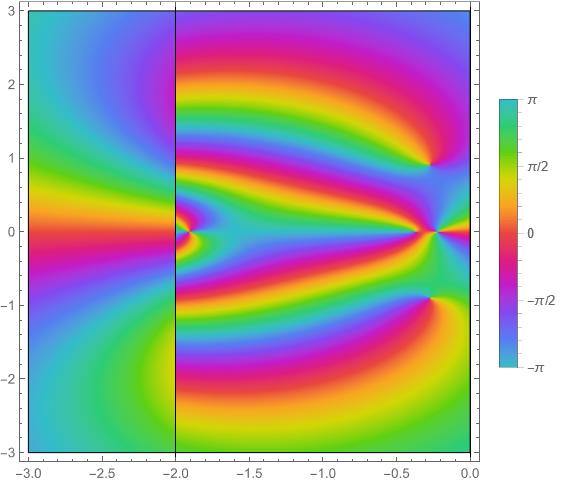}
         \caption{k=0.7}
     \end{subfigure}
        \caption{Argument plot of the spectral function \eqref{defSigma} for relaxation time $\tau = 0.5$, Prandtl number $Pr=0.6$ and different wave numbers $k$. The zeros of \eqref{defSigma} define eigenvalues of the linearized Shakhov mode (points where a small, counter-clockwise loop runs through the whole rainbow at least once). All eigenvalues have negative real part and are located above the essential spectrum $\{\Re\lambda = -\frac{1}{\tau}\}$ (solid black line), which is consistent with the decay estimates \eqref{lowerbound}.\\
        Already at small wave numbers ($k=0.4$ and $k=0.5$), the secondary diffusion and secondary shear mode are close to the essential spectrum ($\{\Re\lambda = -2\}$). For this Prandtl number, the secondary diffusion mode is smaller than the secondary shear mode. At larger wave numbers ($k=0.5$ and $k=0.6$), these modes decrease. At $k=0.7$, the secondary diffusion mode has already disappeared and no merging of branches can occur at this Prandtl number.}
        \label{specplot33}
\end{figure}

In this section, we discuss the phenomenon of branch merging already indicated by \eqref{mode1} and \eqref{mode2} in more details. Through this section, we assume $0\leq r\leq 1$ (for a discussion of $r<0$ and the existence of ghost modes, we refer to the following section). \\
In general, any zero of the spectral function can be expanded in a Puiseux--Newton series with appropriately chosen exponent as already noted in, e.g., \cite{newton1736method}. Namely, the shear modes defined by $\tilde{\Sigma}_1$ can be expanded in a Puiseux--Newton series with exponent $\frac{1}{4}$ (four-fold degeneracy for $k=0$ and $r<1$). The other modes defined by $\tilde{\Sigma}_2$ can be expanded in a Puiseux--Newton series with exponent $\frac{1}{4}$ as well (also a four-fold degeneracy for $k=0$ and $r<1$). A lengthy (and cumbersome) expansion calculation shows, however, that each branch is actually analytic in $k$ and we can therefore expand
\begin{equation}\label{expandlambda}
    \lambda(k,r) = \sum_{k=0}^\infty \lambda_n(r) k^n, 
\end{equation}
where $\lambda_0$ and $\lambda_1$ determine the branch of eigenvalues. This is consistent with the asymptotic expansions for the hydrodynamic branches derived in \cite{ellis1975first} for general kinetic equations and small wave number. Indeed, the instantaneous directional motion of an eigenvalue $k\mapsto\lambda(k)$ is given by
\begin{equation}\label{derlambda}
    \frac{\partial \lambda}{\partial k} = -\frac{\partial_{k}\Sigma_{\mathbf{k},\tau}(\lambda)}{\partial_{\lambda}\Sigma_{\mathbf{k},\tau}(\lambda)},
\end{equation}
whenever $k\mapsto\lambda(k)$ is differentiable.\\
To obtain the coefficients $\lambda_n(r)$, we plug \eqref{expandlambda} into the spectral function \eqref{defSigma} and compare powers of $k$ (using the asymptotic expansion \eqref{Ipsymptotic}):
\begin{equation}
    \Gamma_{|\mathbf{k}|,\tau}\left(\sum_{k=0}^\infty \lambda_n(r) k^n\right) = \sum_{n=0}^\infty G_nk^n.
\end{equation}
Then, we can solve the equations $G_n$ for $\lambda_n$ successively. A lengthy calculation shows that
\begin{equation}
    \lambda_0 \in \left\{\frac{r-1}{\tau}, 0\right\},\quad r<1,
\end{equation}
which is consistent with \eqref{sigmaL0}. To ease notation, we define the two families of branches as
\begin{equation}
\begin{split}
   \lambda_{n,fast}(r) & := \lambda_n\left(r\Big|\lambda_0 =  \frac{r-1}{\tau} \right),\\
   \lambda_{n,slow}(r) & := \lambda_n\left(r|\lambda_0 =  0 \right).
   \end{split}
\end{equation}
Expanding further reveals that
\begin{equation}
    \lambda_{1,fast} = 0,\quad r<1,
\end{equation}
while
\begin{equation}
    \lambda_{1,slow}  \in \left\{0,\pm\ri\sqrt{\frac{5}{3}}\right\},\quad r<1.
\end{equation}
Furthermore, we find that
\begin{equation}\label{l2fast}
    \lambda_{2,fast}(r) = \frac{(81 r-56) \tau }{15 (1-r) r},
\end{equation}
which implies that the secondary diffusion mode moves to the right for $\frac{56}{81}<r<1$, while it moves to the left for $0<r<\frac{56}{81}$ initially. 
\begin{figure}
    \centering
    \includegraphics[width=0.7\textwidth]{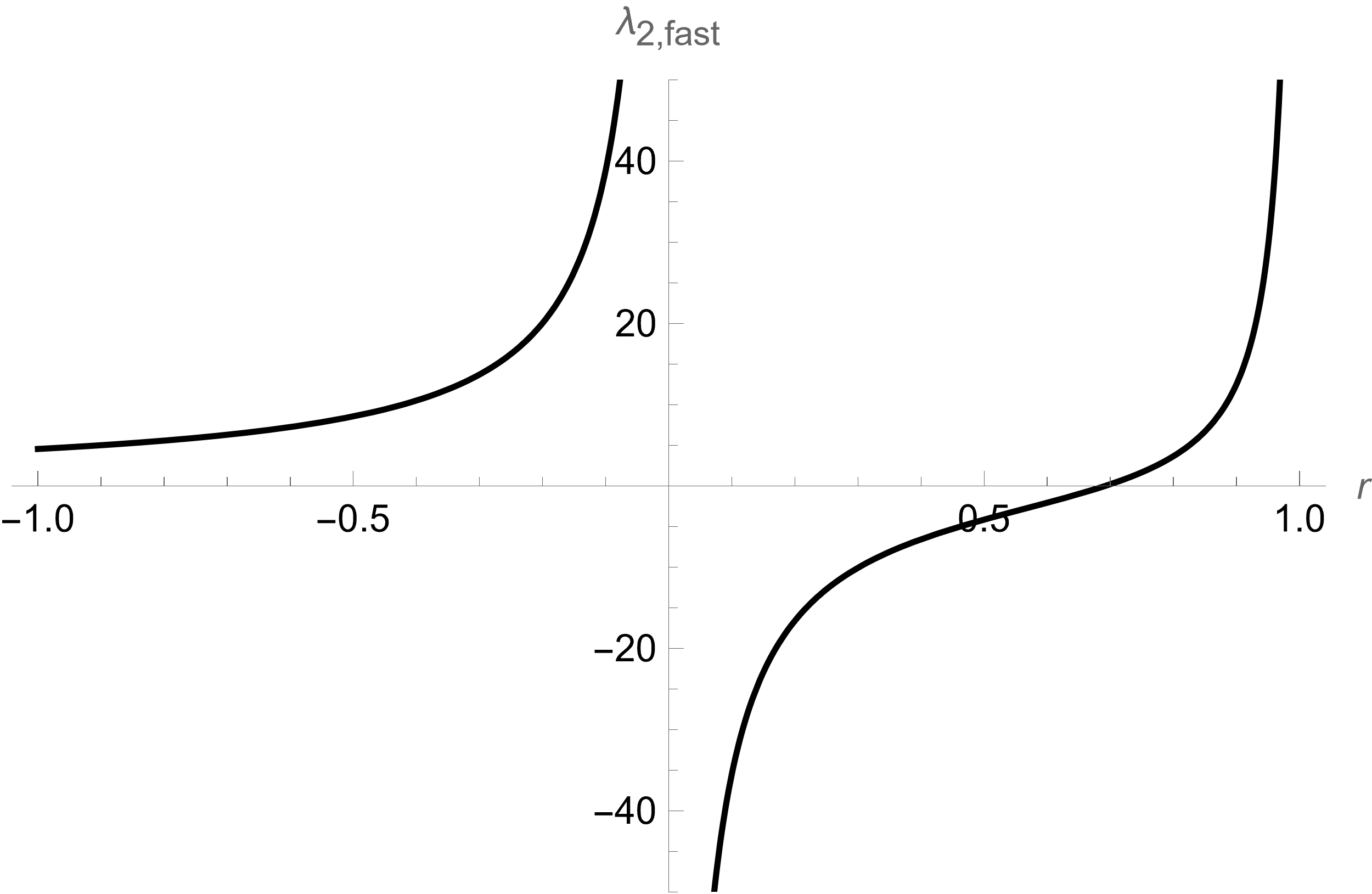}
    \caption{The instantaneous second-order directional motion at $k=0$ of the secondary diffusion mode in dependence of the parameter $r$. For $\frac{56}{81}<r<1$, the secondary diffusion mode moves towards zero until it collides with the primary diffusion mode (second sound). For $0<r<\frac{56}{81}$, the diffusion mode moves towards the essential spectrum at $k=0$ - even with a singularity at $r=0$ (BGK equation). For $r<0$, the secondary diffusion mode is a ghost mode (below the essential spectrum) and always moves towards it instantaneously.}
    \label{l2fastplot}
\end{figure}
This shows that, for $r$ sufficiently close to one, the primary and secondary diffusion branch will inevitably collide and produce a pair of complex-conjugate zeros (secondary acoustics) through a saddle-node bifurcation. \\
On the other hand, we see that for $r$ close to $0$, the secondary diffusion mode will travel to the left from the very beginning and will leaf the domain $-\frac{1}{\tau}<\Re\lambda<0$ before it gets the chance to collide with the primary diffusion branch, see Figure \ref{specplot33}. Consequently, there does not exist the phenomenon of branch merging and second sound. Indeed, in the limit $r\to 0$, the Shakhov S-model reduces to the BGK model, for which no second sound exists. We summarize that - if the Prandtl number is close to one, i.e., close to the dynamics of the three-dimensional linear BGK equation, there exists no second sound (clearly, there is no second sound for the BGK equation), while for Prandtl number close to zero, there will always be branch merging.



\begin{remark}
While the coefficient \eqref{l2fast} governs the motion of the secondary diffusion mode for small wave numbers, the behavior for larger $k$ might be different. Indeed, for a certain value of $r$, the secondary diffusion mode might start out by moving to the left, but then turn to the right and collide with primary diffusion mode anyways. The above considerations thus only imply that there is a certain range for which second sound exists and that there is a certain range for which it does not exist.
\end{remark}

\subsection{Spectral Properties for Prandtl number greater than one: existence of ghost modes}\label{spectral5}

\begin{figure}
     \centering
     \begin{subfigure}[b]{0.45\textwidth}
         \centering
         \includegraphics[width=\textwidth]{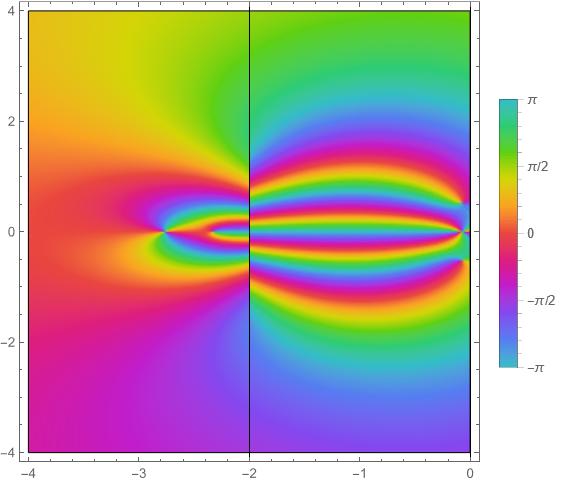}
         \caption{$k=0.4$}
     \end{subfigure}
     \begin{subfigure}[b]{0.45\textwidth}
         \centering
         \includegraphics[width=\textwidth]{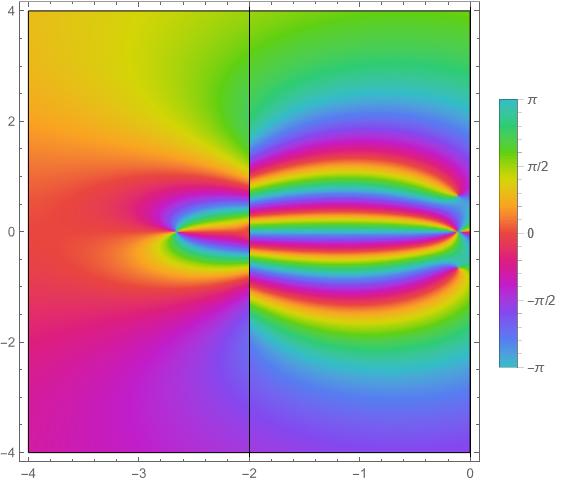}
         \caption{k=0.5}
     \end{subfigure}
     \begin{subfigure}[b]{0.45\textwidth}
         \centering
         \includegraphics[width=\textwidth]{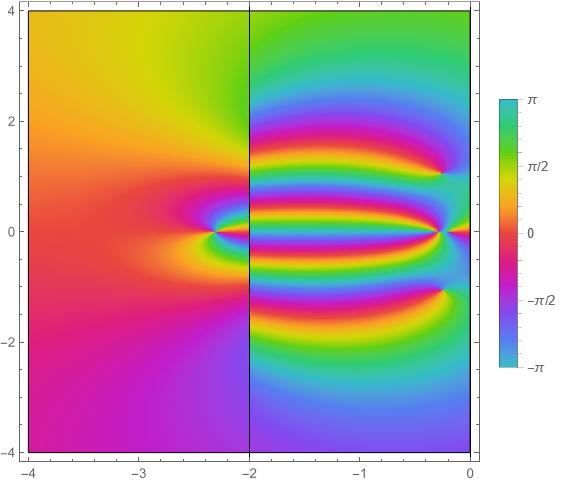}
         \caption{k=0.8}
     \end{subfigure}
     \begin{subfigure}[b]{0.45\textwidth}
         \centering
         \includegraphics[width=\textwidth]{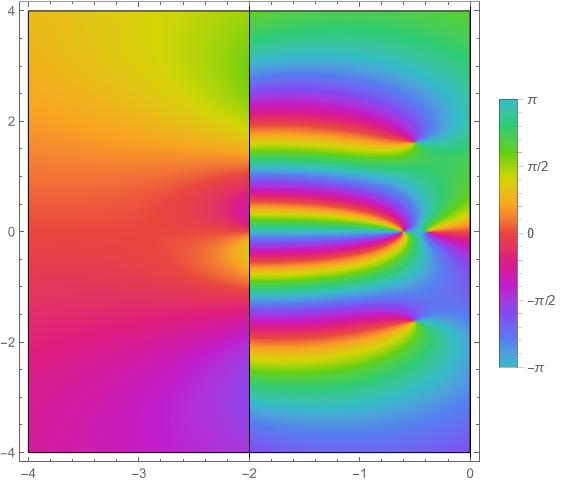}
         \caption{k=1.2}
     \end{subfigure}
        \caption{Argument plot of the spectral function \eqref{defSigma} for relaxation time $\tau = 0.5$, Prandtl number $Pr=1.5$ ($r=-0.5$) and different wave numbers $k$. The zeros of \eqref{defSigma} define eigenvalues of the linearized Shakhov mode (points where a small, counter-clockwise loop runs through the whole rainbow at least once). All eigenvalues have negative real part, but some of them are located below the essential spectrum $\{\Re\lambda = -\frac{1}{\tau}\}$ (solid black line), which is consistent with the decay estimates \eqref{lowerboundnegative}.\\
        At small wave numbers, three ghost modes, namely a shear mode and a diffusion mode, appear below the essential spectrum ($k=0.4$). As the wave number is increased, the ghost modes move closer towards the essential spectrum until the diffusion mode is absorbed ($k=0.5$, $k=0.8$). Finally, also the ghost shear mode is absorbed ($k=1.2$) and the spectrum consist of primary hydrodynamic modes above the essential spectrum only (up to the critical wave number).}
        \label{rneg}
\end{figure}

In this section, we derive the behavior of the spectrum of \eqref{linmain} for $r<0$ (or $Pr>1$). 
As already indicated by the estimate \eqref{lowerboundnegative}, potential eigenvalues are no longer guaranteed to exist above the essentially spectrum exclusively for $r<0$. Figure \eqref{rneg} shows some typical argument plots of the spectral function for $Pr=1.5$ ($r=-0.5$) with eigenvalues below the essential spectrum.\\
Obviously, from the considerations around the spectrum of $\hat{\mathcal{L}}_{0}$ in \eqref{sigmaL0}, we see that for $Pr>1$ (which is equivalent to $\tau_{\rm fast}>\tau_{\rm slow}$), the eigenvalue $-\frac{1}{\tau_{\rm slow}}$ is indeed located below the essential spectrum $\{\Re\lambda = -\frac{1}{\tau_{\rm fast}}\}$ : 
\begin{equation}
    -\frac{1}{\tau_{\rm slow}}<-\frac{1}{\tau_{\rm fast}}. 
\end{equation}
From the instantaneous directional motion of the secondary diffusion mode \eqref{l2fast}, we see that
\begin{equation}
    \lambda_{2,fast}(r) = \frac{(81 r-56) \tau }{15 (1-r) r}>0,\quad r<0,
\end{equation}
which shows that the ghost (diffusion) mode increases with wave number until it is absorbed by the essential spectrum, see Figure \ref{rneg}. A similar argument holds true for the secondary shear modes. 
Existence of non-hydrodynamic modes below the essential spectrum suggests that Shakhov's model becomes unrealistic  $\rm Pr>1$, and other kinetic models, such as the ES-BGK should be used instead for $\rm Pr>1$.

\section{Linear Hydrodynamic Manifolds}\label{sechydro}

We define a hydrodynamic manifold as the (unique) slow spectral submanifold associated to a set of eigenvectors. In the case of linear dynamics, the manifold itself is given by the invariant linear subspace spanned by the eigenvalues associated to a set of slow eigenvalues. In particular, the hydrodynamic manifold has the following properties:
\begin{enumerate}
\item It contains an appropriately scaled, spatially independent stationary distribution (e.g. global Maxwellian) as a base solution
\item The projection onto the hydrodynamic moments along the manifold provide a closure of the hydrodynamic moments (mass-density, velocity and temperature)
\item It attracts \textit{all} trajectories in the space of probability-density functions (which are close enough to the base solution) exponentially fast, thus acting as a slow manifold
\item It is unique. 
\end{enumerate}

We write the hydrodynamic manifold in frequency space as a superposition of eigen-functions associated to each mode 
\begin{equation}
\hat{f}_{\rm hydro}(\mathbf{k},\mathbf{v},t) = \sum_{n\in \text{Modes}(k)} \beta(t,\mathbf{k})\hat{f}^{eig}_n(\mathbf{k},\mathbf{v}),
\end{equation}
where the set of modes is given by $\text{Modes}=\{\rm ac_1,ac_1*, diff_1, shear_1, diff_2, shear_2\}$ or  $\text{Modes}=\{\rm ac_1,ac_1*, shear_1, ac_2,ac_2*,shear_2\}$ depending on $k$ and $r$. To omit cluttering of the notation, we omit the second elements in the Jordan block generated by the shear
The frequency-dependent eigen-functions solve the equation
\begin{equation}
-\ri\mathbf{k}\cdot\mathbf{v}\hat{f}^{eig}_n -\frac{1}{\tau}\hat{f}^{eig}_n+\mathbb{B}_{8,r} \hat{f}^{eig}_n = \lambda_n \hat{f}^{eig}_n
\end{equation}
\begin{equation}\label{eigf2ts}
\hat{f}^{eig}_n(\mathbf{k},\mathbf{v}) = \frac{\mathbf{e}(\mathbf{v})\cdot\boldsymbol{\alpha}_{n}}{\tau\ri\mathbf{k}\cdot\mathbf{v}+1+\tau\lambda_n(k\tau)},
\end{equation}
which satisfies
\begin{equation}\label{alpha}
    \boldsymbol{\alpha}_n = \langle  \hat{f}^{eig}_n,\mathbf{e}\rangle_{\mathbf{v}}.
\end{equation}
Indeed, equation \eqref{alpha} is equivalent to 
\begin{equation}
    \boldsymbol{\alpha}_n \in \text{ker}(\mathbf{D}_rG_S-Id)_{z=-1-\tau\lambda_n},
\end{equation}
for the matrix
\begin{equation}
    G_S(z)=\int_{\mathbb{R}^3} \mathbf{e}(\mathbf{v})\otimes\mathbf{e}(\mathbf{v})\frac{e^{-\frac{|\mathbf{v}|}{2}}}{\tau\ri\mathbf{k}\cdot\mathbf{v}-z} \, d\mathbf{v}
\end{equation}
We define the hydrodynamic variables as
\begin{equation}\label{defhdefHBGK}
    \hat{\mathbf{h}}=\langle f_{\rm hydro},\mathbf{e}\rangle_{\mathbf{v}},
\end{equation}
which gives
\begin{equation}\label{Hhat}
   \hat{\mathbf{h}} = \langle f_{\rm hydro},\mathbf{e}\rangle_{\mathbf{v}} = \sum_{n\in M} \beta_n\langle \hat{f}^{eig}_n,\mathbf{e}\rangle_{\mathbf{v}}=\mathbf{A}\boldsymbol{\beta},
\end{equation}
where, depending on $r$ and $k$, either
\begin{equation}
    \mathbf{A}=(\boldsymbol{\alpha}_{\rm ac1},\boldsymbol{\alpha}_{\rm ac1*},\boldsymbol{\alpha}_{\rm diff1},\boldsymbol{\alpha}_{\rm shear1},\boldsymbol{\alpha}_{\rm diff2},\boldsymbol{\alpha}_{\rm shear2}),
\end{equation}
or 
\begin{equation}
    \mathbf{A}=(\boldsymbol{\alpha}_{\rm ac1},\boldsymbol{\alpha}_{\rm ac1*},\boldsymbol{\alpha}_{\rm shear1},\boldsymbol{\alpha}_{\rm ac2},\boldsymbol{\alpha}_{\rm ac2*}\boldsymbol{\alpha}_{\rm shear2}).
\end{equation}
The vector $\boldsymbol{\beta}$ describes the evolution of the hydrodynamic variables in terms of the basis of eigenvalues (spectral basis). The time-evolution of the hydrodynamics can then be written as
\begin{equation}
\hat{\mathbf{h}}_t=\langle \partial_tf_{\rm hydro},\mathbf{e}\rangle_{\mathbf{v}} = \sum_{n\in \text{Modes}(k)}\langle \hat{f}^{eig}_n,\mathbf{e}\rangle_{\mathbf{v}}\lambda_n\beta_n=\Theta \mathbf{A}\Lambda\boldsymbol{\beta},
\end{equation}
or, solving for $\beta$ in  \eqref{Hhat}:
\begin{equation}\label{HdynBGK}
\hat{\mathbf{h}}_t = \mathbf{A}\Lambda \mathbf{A}^{-1}\hat{\mathbf{h}}.
\end{equation}
Equation \eqref{HdynBGK} defines the exact hydrodynamics derived from the slow motion along the hydrodynamic manifold. 

\begin{remark}
The evaluation of the right-hand side of \eqref{HdynBGK} is by no means trivial and involves properties of the spectral projection. The properties of the exact, spectrally-closed hydrodynamics together with their physical properties will be discussed in a forthcoming paper \cite{spectralclosure}. 
\end{remark}


\section{Conclusion and Further Perspectives}

We performed a complete spectral analysis for the Shakhov model linearized around a global Maxwellian. The discrete eigenvalues above the essential spectrum $\{\Re\lambda = -\frac{1}{\tau_{\rm fast}}\}$ are described as zeros of a spectral function at each wave number. In this way, we identified families of modes (branches), depending on wave number. For small wave numbers, the family of modes is given by $\text{Modes} = \{\rm ac_{1},ac_1*, diff_1, diff_2, shear_1, shear_2\}$, the pair of primary acoustic modes, the primary and secondary diffusion
modes as well as the primary and secondary shear modes. Within a certain range of Prandtl numbers, a merging of branches can occur at a specific wave number and the modes $\rm diff_1$ and $\rm diff_2$ may collide, producing another pair of acoustic modes $\rm ac_2$ and $\rm ac_2*$ via a saddle-node bifurcation. This phenomenon is known as \textit{second sound}.\\
The approach presented in \cite{kogelbauer2021,kogelbauer2023exact} as well as in the current paper is general enough to infer spectral properties for any finitely-truncated collision operator, such as quasi-equilibrium approximations \cite{GORBAN2006325} or even Maxwell molecules \cite{truesdell1980fundamentals}. The explicit knowledge and quantitative properties of the spectra identified for several kinetic model equations \cite{kogelbauer2021,kogelbauer2023exact} also allow us to move to the existence theory of non-linear hydrodynamic equations for various (finitely-truncated) kinetic models. Indeed, the fact that the discrete spectrum is well separated from the essential spectrum allows us to define a spectral projection for the \textit{whole} set of eigenvalues, thus giving the first-order approximation (in terms of nonlinear deformations) to the hydrodynamic manifolds. In particular, we expect that the theory of thermodynamic projectors \cite{GORBAN2004391} may be helpful in proving the nonlinear extension.\\
The quantitative insights in the structure of the spectrum could also be used to derive simplified, but still non-local, approximate hydrodynamics. This could also improve present numerical methods
\cite{karlin2008exact}.\\

\section*{Acknowledgement}
This work was supported by European Research Council (ERC) Advanced Grant 834763-PonD. Computational
resources at the Swiss National Super Computing Center CSCS were provided under the grant s1066.

\section*{Declaration of Interest}
The authors declare that there is no conflict of interests.

\appendix

\section{Linearization and Non-Dimensionalization of the Shakhov Equation Around a Global Maxwellian}
In this section we perform - for the sake of completeness - the linearization of the Shakhov model around a global Maxwellian. To obtain the linearization of \eqref{maineq} around $F_0^{eq}$, we write $F = F_0^{eq}+\varepsilon f$ and calculate
\begin{equation}\label{lin1}
\begin{split}
\left. \frac{d}{d\varepsilon}\right|_{\varepsilon=0} Q_{fs}[F_0^{eq}+\varepsilon f]&=f-\left(\left.\frac{d}{d\varepsilon}\right|_{\varepsilon=0}F^{eq}[F_0^{eq}+\varepsilon f]\right)\left(1+(1-Pr)\frac{\mathbf{q}[F_0^{eq}]\cdot(\mathbf{v}-\mathbf{u}[F_0^{eq}])}{Rp[F_0^{eq}]T[F_0^{eq}]}\left(\frac{|\mathbf{v}-\mathbf{u}[F_0^{eq}]|^2}{5RT[F_0^{eq}]}-1\right)\right)\\
&\qquad-F^{eq}[F_0^{eq}]r\left.\frac{d}{d\varepsilon}\right|_{\varepsilon=0}\frac{\mathbf{q}[F_0^{eq}]\cdot(\mathbf{v}-\mathbf{u}[F_0^{eq}+\varepsilon f])}{Rp[F_0^{eq}+\varepsilon f]T[F_0^{eq}+\varepsilon f]}\left(\frac{|\mathbf{v}-\mathbf{u}[F_0^{eq}+\varepsilon f]|^2}{5RT[F_0^{eq}+\varepsilon f]}-1\right)
\end{split}
\end{equation}
Using the relations
\begin{equation}
\begin{split}
n[F^{eq}_0]= n_0,\quad\mathbf{u}[F^{eq}_0] = 0,\quad\mathbf{q}[F_0^{eq}]=0,\quad p[F^{eq}_0] = mRn_0T_0,\quad T[F_0^{eq}] = T_0,
\end{split}
\end{equation}
which in particular imply that $F^{eq}[F_0^{eq}] = F_0^{eq}$, we can reduce \eqref{lin1} to
\begin{equation}\label{lin2}
\begin{split}
\left. \frac{d}{d\varepsilon}\right|_{\varepsilon=0} Q_{fs}[F_0^{eq}+\varepsilon f]&=f-\left(\left.\frac{d}{d\varepsilon}\right|_{\varepsilon=0}F^{eq}[F_0^{eq}+\varepsilon f]\right)\\
&\qquad-rF_0^{eq}\left.\frac{d}{d\varepsilon}\right|_{\varepsilon=0}\frac{\mathbf{q}[F_0^{eq}+\varepsilon f]\cdot(\mathbf{v}-\mathbf{u}[F_0^{eq}+\varepsilon f])}{Rp[F_0^{eq}+\varepsilon f]T[F_0^{eq}+\varepsilon f]}\left(\frac{|\mathbf{v}-\mathbf{u}[F_0^{eq}+\varepsilon f]|^2}{5RT[F_0^{eq}+\varepsilon f]}-1\right).
\end{split}
\end{equation}
Denoting
\begin{equation}\label{defm}
    \mathbf{m}_n = \int_{\mathbf{R}^3} f(\mathbf{x},\mathbf{v},t) \mathbf{v}^{\otimes n}\, d\mathbf{v},
\end{equation}
the moments of the perturbation density $f$, the moments \eqref{defmoment} transform according to
\begin{equation}
\begin{split}
\mathbf{M}_0&= n_0 +\varepsilon\mathbf{m}_0,\\
\mathbf{M}_1&= \varepsilon\mathbf{m}_1,\\
\mathbf{M}_2&= Rn_0T_0 \Id_{3\times3} + \varepsilon\mathbf{m}_2,\\
\mathbf{M}_3&= \varepsilon \mathbf{m}_3,
\end{split}
\end{equation}
which in turn implies that
\begin{equation}
    \begin{split}
        n & =n_0 + \varepsilon\mathbf{m}_0 ,\\
      \mathbf{u} & = \frac{\varepsilon\mathbf{m}_1}{n_0+\varepsilon\mathbf{m}_0} ,\\
      p & = \frac{m}{3}(3Rn_0T_0+\varepsilon\tr\mathbf{m}_3)-\varepsilon^2\frac{m}{3}\frac{|\mathbf{m}_1|^2}{n_0 + \varepsilon\mathbf{m}_0},\\
      \mathbf{q} & = -\frac{3}{2}\left(\frac{m}{3}(3Rn_0T_0+\varepsilon\tr\mathbf{m}_3)-\varepsilon^2\frac{m}{3}\frac{|\mathbf{m}_1|^2}{n_0 + \varepsilon\mathbf{m}_0}\right)\frac{\varepsilon\mathbf{m}_1}{n_0 + \varepsilon\mathbf{m}_0}\\
      &\qquad+\varepsilon\frac{m}{2}\tilde{\mathbf{m}}_3+\frac{m}{2}\frac{\varepsilon^3|\mathbf{m}_1|^2}{(n_0+\varepsilon\mathbf{m}_0)^3}\mathbf{m}_1-m(Rn_0T_0 \Id_{3\times3} + \varepsilon\mathbf{m}_2)\frac{\varepsilon\mathbf{m}_1}{n_0+\varepsilon\mathbf{m}_0}.
    \end{split}
\end{equation}
Consequently, the $\varepsilon$-derivatives of the hydrodynamic moments become
\begin{equation}\label{linnuTq}
\begin{split}
\left.\frac{\partial n}{\partial \varepsilon}\right|_{\varepsilon=0}&=\mathbf{m}_0,\\
\left.\frac{\partial \mathbf{u}}{\partial \varepsilon}\right|_{\varepsilon=0}&=\frac{\mathbf{m}_1}{n_0},\\
\left.\frac{\partial p}{\partial \varepsilon}\right|_{\varepsilon=0}&=\frac{m}-{3}\tr\mathbf{m}_2,\\
\left.\frac{\partial q}{\partial \varepsilon}\right|_{\varepsilon=0}&=\frac{3mRT_0}{2}\mathbf{m}_1+\frac{m}{2}\tilde{\mathbf{m}}_3-mRT_0\mathbf{m}_1=-\frac{5mRT_0}{2}\mathbf{m}_1+\frac{m}{2}\tilde{\mathbf{m}}_3.
\end{split}
\end{equation}
With \eqref{linnuTq},  we can calculate
\begin{equation}\label{dereq}
\begin{split}
\left.\frac{\partial}{\partial \varepsilon}\right|_{\varepsilon=0} F^{eq}[F_{0}^{eq} +\varepsilon f]&=\left(2\pi RT_0\right)^{-\frac{3}{2}}e^{-\frac{|\mathbf{v}|^2}{2RT_0}}\left(\mathbf{m}_0-\frac{\tr\mathbf{m}_2-3RT_0\mathbf{m}_0}{2RT_0}\right.\\
&\qquad \left.+\frac{\mathbf{m}_1\cdot\mathbf{v}}{RT_0}+\frac{\tr\mathbf{m}_2-3RT_0\mathbf{m}_0}{6}\frac{|\mathbf{v}|^2}{(RT_0)^2}\right),
\end{split}
\end{equation}
while
\begin{equation}\label{derq}
\begin{split}
    F_0^{eq}[F_{0}^{eq}]\left.\frac{d}{d\varepsilon}\right|_{\varepsilon=0}&\frac{\mathbf{q}[F_0^{eq}+\varepsilon f]\cdot(\mathbf{v}-\mathbf{u}[F_0^{eq}+\varepsilon f])}{Rp[F_0^{eq}+\varepsilon f]T[F_0^{eq}+\varepsilon f]}\left(\frac{|\mathbf{v}-\mathbf{u}[F_0^{eq}+\varepsilon f]|^2}{5RT[F_0^{eq}+\varepsilon f]}-1\right)  \\
    &\qquad=n_0\left(2\pi RT_0\right)^{-\frac{3}{2}}e^{-\frac{|\mathbf{v}|^2}{2RT_0}}\left(\frac{|\mathbf{v}|^2}{5RT_0}-1\right)\frac{1}{Rp_0T_0}\left.\frac{d\mathbf{q}[F_0^{eq}+\varepsilon f]}{d\varepsilon}\right|_{\varepsilon=0}\cdot \mathbf{v}\\
    &\qquad=\left(2\pi RT_0\right)^{-\frac{3}{2}}e^{-\frac{|\mathbf{v}|^2}{2RT_0}}\left(\frac{|\mathbf{v}|^2}{5RT_0}-1\right)\frac{n_0}{Rp_0T_0}\left(-\frac{5mRT_0}{2}\mathbf{m}_1+\frac{m}{2}\tilde{\mathbf{m}}_3\right)\cdot \mathbf{v}.
    \end{split}
\end{equation}
Combining \eqref{dereq} with \eqref{derq}, equation \eqref{lin2} reads
\begin{equation}
\begin{split}
\left. \frac{d}{d\varepsilon}\right|_{\varepsilon=0} Q_{fs}[F_0^{eq}+\varepsilon f]&=f-\left(2\pi RT_0\right)^{-\frac{3}{2}}e^{-\frac{|\mathbf{v}|^2}{2RT_0}}\left(\mathbf{m}_0-\frac{\tr\mathbf{m}_2-3RT_0\mathbf{m}_0}{2RT_0}+\frac{\mathbf{m}_1\cdot\mathbf{v}}{RT_0}\right.\\
&\qquad\left.+\frac{\tr\mathbf{m}_2-3RT_0\mathbf{m}_0}{6}\frac{|\mathbf{v}|^2}{(RT_0)^2}+r\left(\frac{|\mathbf{v}|^2}{5RT_0}-1\right)\frac{1}{(RT_0)^2}\left(-\frac{5RT_0}{2}\mathbf{m}_1+\frac{1}{2}\tilde{\mathbf{m}}_3\right)\cdot \mathbf{v}\right)
\end{split}
\end{equation}
Defining the \textit{thermal velocity} as
\begin{equation}\label{defvthermal}
v_{thermal}=\sqrt{RT_0},
\end{equation}
and re-scaling according to
\begin{equation}
\mathbf{v}\mapsto v_{thermal}\mathbf{v},
\end{equation}
implies that
\begin{equation}
\mathbf{m}_n\mapsto \left(RT_0\right)^{\frac{3+n}{2}}\mathbf{m}_n,
\end{equation}
which allows us to simplify
\begin{equation}
\begin{split}
\left.\frac{\partial}{\partial \varepsilon}\right|_{\varepsilon=0} Q_{eq}[F_0^{eq} +\varepsilon f]
=f-(2\pi)^{-3/2}e^{-\frac{|\mathbf{v}|^2}{2}}\left(\frac{5\mathbf{m}_0-\tr\mathbf{m}_2}{2}+\mathbf{m}_1\cdot\mathbf{v}\right.\\
\left.+\frac{\tr\mathbf{m}_2-3\mathbf{m}_0}{6}|\mathbf{v}|^2+r\left(\frac{|\mathbf{v}|^2}{5}-1\right)\left(\frac{\tilde{\mathbf{m}}_3-5\mathbf{m}_1}{2}\right)\cdot \mathbf{v}\right)\\
=f-(2\pi)^{-3/2}e^{-\frac{|\mathbf{v}|^2}{2}}\left[\left(\frac{5-|\mathbf{v}|^2}{2}\right)\mathbf{m}_0+\left(1-\frac{5r}{2}\left(\frac{|\mathbf{v}|^2}{5}-1\right)\right)\mathbf{v}\cdot\mathbf{m}_1\right.\\
\left.+\left(\frac{|\mathbf{v}|^2-3}{2}\right)\tr\mathbf{m}_2+r\left(\frac{|\mathbf{v}|^2-5}{10}\right)\mathbf{v}\cdot\tilde{\mathbf{m}}_3\right]
\end{split}
\end{equation}
Similarly, we re-scale
\begin{equation}
\mathbf{x}\mapsto L\mathbf{x},
\end{equation}
which implies that $\mathbf{x}\in\mathbb{T}^3$ henceforth. Defining the thermal time
\begin{equation}
t_{thermal} = L\sqrt{\frac{m}{k_BT_0}},
\end{equation}
we can re-scale and non-dimensionalize
\begin{equation}
t\mapsto t t_{thermal},\qquad \tau \mapsto \tau t_{thermal},
\end{equation}
and finally arrive at the linearized and non-dimensionalized Shakhov equation:
\begin{equation}\label{linmainapp}
    \begin{split}
        \frac{\partial f}{\partial t} =-\mathbf{v}\cdot\nabla_{\mathbf{x}}f-\frac{1}{\tau}f+\frac{1}{\tau} (2\pi)^{-3/2}e^{-\frac{|\mathbf{v}|^2}{2}}\left[\left(\frac{5-|\mathbf{v}|^2}{2}\right)\mathbf{m}_0+\left(1+\frac{r}{2}\left(\frac{|\mathbf{v}|^2-5}{5}\right)\right)\mathbf{v}\cdot\mathbf{m}_1\right.\\
\left.+\left(\frac{|\mathbf{v}|^2-3}{2}\right)\tr\mathbf{m}_2+r\left(\frac{|\mathbf{v}|^2-5}{10}\right)\mathbf{v}\cdot\tilde{\mathbf{m}}_3\right].
    \end{split}
\end{equation}

\section{Properties of the Plasma Dispersion Function $Z$}

\begin{centering}
\begin{figure}
	\begin{subfigure}{.5\textwidth}
		\centering
		\includegraphics[width=.8\linewidth]{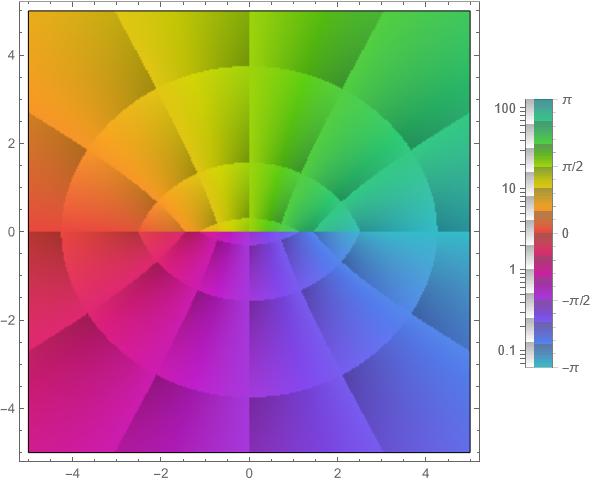}
		\caption{Argument Plot of $Z$}
	\end{subfigure}%
	\begin{subfigure}{.5\textwidth}
		\centering
		\includegraphics[width=1\linewidth]{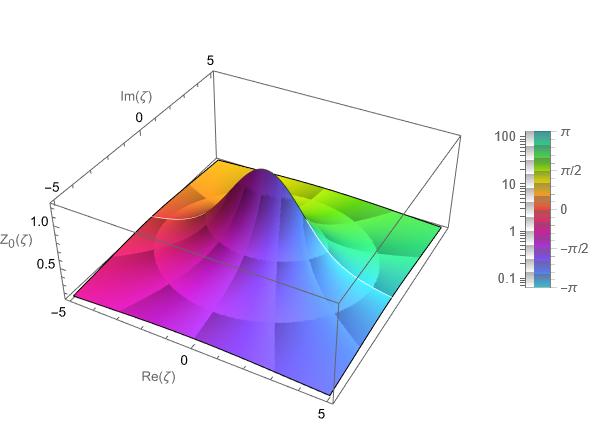}
  \caption{Modulus-Argument Plot of $Z$}
	\end{subfigure}
 \caption{Complex plots of the function Z.}
\label{pI0}
\end{figure}
\end{centering}

In the following, we collect some properties of the plasma dispersion function $Z$, defined through the  integral expression \eqref{defZ}. In our presentation, we will closely follow the calculations performed in \cite{kogelbauer2023exact}.\\
First, let us derive an expression of the integral \eqref{defZ} in terms of less exotic functions. To this end, we rely on the identities in \cite[p.297]{abramowitz1948handbook}. Let
\begin{equation}\label{defw}
w(\zeta)=e^{-\zeta^2}(1-\erf(-\ri \zeta)), \quad \zeta\in\mathbb{C},
\end{equation}
which satisfies the functional identity
\begin{equation}\label{wident}
w(-\zeta)=2e^{-\zeta^2}-w(\zeta),\quad \zeta\in\mathbb{C}.
\end{equation}
Function \eqref{defw} is called \textit{Faddeeva function} and is frequently encountered in problems related to kinetic equations \cite{fitzpatrick2014plasma}. 
We then have that
\begin{equation}
w(\zeta)=\frac{\ri}{\pi}\int_{\mathbb{R}}\frac{e^{-s^2}}{\zeta-s}\, ds,\quad \Im{\zeta}>0,
\end{equation}
and, by relation \eqref{wident}, we have for $\Im{\zeta}<0$:
\begin{equation}
\begin{split}
\frac{\ri}{\pi}\int_{\mathbb{R}}\frac{e^{-s^2}}{\zeta-s}\, ds&=-\frac{\ri}{\pi}\int_{\mathbb{R}}\frac{e^{-s^2}}{(-\zeta)+s}\, ds\\
&=-\frac{\ri}{\pi}\int_{\mathbb{R}}\frac{e^{-s^2}}{(-\zeta)-s}\, ds\\
&=-w(-\zeta)\\
&=e^{-\zeta^2}[-1-\erf(-\ri \zeta)].
\end{split}
\end{equation}
Consequently, we obtain
\begin{equation}
\begin{split}
\int_{\mathbb{R}}\frac{1}{s-\zeta}e^{-\frac{s^2}{2}}\, ds&=\int_{\mathbb{R}}\frac{e^{-s^2}}{s-\frac{\zeta}{\sqrt{2}}}\, ds\\
&=\ri\pi\frac{\ri}{\pi}\int_{\mathbb{R}}\frac{e^{-s^2}}{\frac{\zeta}{\sqrt{2}}-s}\, ds\\
&=\begin{cases}
\ri\pi e^{-\frac{\zeta^2}{2}}\left[1-\erf\left(\frac{-\ri \zeta}{\sqrt{2}}\right)\right],&\quad\text{ if } \Im{\zeta}>0,\\
\ri \pi e^{-\frac{\zeta^2}{2}}\left[-1-\erf\left(\frac{-\ri \zeta}{\sqrt{2}}\right)\right],&\quad\text{ if } \Im{\zeta}<0,
\end{cases}
\end{split}
\end{equation}
where in the first step, we have re-scaled $s\mapsto \sqrt{2}s$ in the integral.
Written more compactly, we arrive at
\begin{equation}\label{g0}
Z(\zeta)=\ri\sqrt{\frac{\pi}{2}} e^{-\frac{\zeta^2}{2}}\left[\sign(\Im{\zeta})-\erf\left(\frac{-\ri \zeta}{\sqrt{2}}\right)\right], \quad \Im{\zeta}\neq 0. 
\end{equation}
An an argument plot together with an modulus-argument plot of $Z$ are shown in Figure \ref{pI0}.
Clearly, $Z$ is discontinuous across the real line (albeit that $Z|_{\mathbb{R}}$ exists in the sense of principal values as the Hilbert transform of a real Gaussian \cite{conte1961plasma}). The properties 
\begin{equation}
\begin{split}
    |Z(\zeta)|\leq   \sqrt{\frac{\pi}{2}}, &\text{ for } \zeta\in\mathbb{C}\setminus\mathbb{R},\\
    0< \arg Z(\zeta)<\pi &\text{ for } \Im(\zeta)>0,\\
    -\pi < \arg Z(\zeta) <0 &\text{ for } \Im(\zeta)<0,
\end{split}
\end{equation}
are easy to show and can be read off from the plots \eqref{pI0} directly as well.\\
We also note that
\begin{equation}\label{Z0}
\begin{split}
\lim_{\zeta\to 0,\Im\zeta>0} Z(\zeta) = \ri\sqrt{\frac{\pi}{2}},\\
\lim_{\zeta\to 0,\Im\zeta<0} Z(\zeta) = -\ri\sqrt{\frac{\pi}{2}},
\end{split}
\end{equation}
as can be seen from \eqref{g0}.\\
Function \eqref{g0} satisfies an ordinary differential equation (in the sense of complex analytic functions) on the upper and on the lower half-plane. Indeed, integrating \eqref{defZ} by parts gives
\begin{equation}
\begin{split}
1 &= \frac{1}{\sqrt{2\pi}} \int_{\mathbb{R}}(v-\zeta)\frac{e^{-\frac{v^2}{2}}}{v-\zeta}\, dv=-\zeta Z+\frac{1}{\sqrt{2\pi}}\int_{\mathbb{R}}v\frac{e^{-\frac{v^2}{2}}}{v-\zeta}\, dv\\
&=-\zeta Z-\frac{1}{\sqrt{2\pi}}\int_{\mathbb{R}}\frac{e^{-\frac{v^2}{2}}}{(v-\zeta)^2}\, dv=-\zeta Z-\frac{d}{d\zeta}Z,
\end{split}
\end{equation} 
which implies that $Z$ satisfies the differential equation
\begin{equation}\label{dI0}
\frac{d}{d\zeta}Z= -\zeta Z-1,
\end{equation}
for $\zeta\in\mathbb{C}\setminus\mathbb{R}$. Formula \eqref{dI0} can also be used as a recurrence relation for the higher derivatives of $Z$.\\
Since we will be interested in function \eqref{g0} for $\Im \zeta$ positive and negative as global functions, we define
\begin{equation}\label{defIplus}
\begin{split}
    Z_{+}(\zeta) &= \ri\sqrt{\frac{\pi}{2}} e^{-\frac{\zeta^2}{2}}\left[1-\erf\left(\frac{-\ri \zeta}{\sqrt{2}}\right)\right],\\
    Z_{-}(\zeta) &= \ri\sqrt{\frac{\pi}{2}} e^{-\frac{\zeta^2}{2}}\left[-1-\erf\left(\frac{-\ri \zeta}{\sqrt{2}}\right)\right],
    \end{split}
\end{equation}
for all $\zeta\in\mathbb{C}$. Both functions can be extended to analytic functions on the whole complex plane via analytic continuation.\\
Recall that the error function has the properties that
\begin{equation}
\erf(-\zeta)=-\erf(\zeta),\qquad \erf(\zeta^*)=\erf(\zeta)^*,
\end{equation}
for all $\zeta\in\mathbb{C}$, which implies that for $x\in\mathbb{R}$,
\begin{equation}\label{erfi}
    \erf(\ri x)=-\erf(-\ri x)=-\erf(\ri x)^*,
\end{equation}
i.e, the error function maps imaginary numbers to imaginary numbers. Defining the \textit{imaginary error function},
\begin{equation}
\erfi(\zeta):=-\ri \erf(\ri \zeta),
\end{equation}
for $\zeta\in\mathbb{C}$, which, by \eqref{erfi} satisfies $\erfi|_{\mathbb{R}}\subset\mathbb{R}$, it follows that for $x\in\mathbb{R}$:
\begin{equation}\label{ReImIplus}
  \Re Z_{+}(x)= -\sqrt{\frac{\pi}{2}}e^{-\frac{x^2}{2}}\erfi\left(\frac{x}{\sqrt{2}}\right),\quad \Im Z_{+}(x)= -\sqrt{\frac{\pi}{2}}e^{-\frac{x^2}{2}},
\end{equation}
similarly for $Z_{-}(x)$.\\
Next, let us prove the following asymptotic expansion of $Z_+$:
 \begin{equation}\label{Ipsymptotic}
 Z_{+}(\zeta) \sim -\sum_{n=0}^\infty \frac{(2n-1)!!}{\zeta^{2n+1}},  \qquad \text{ for }|\arg(\zeta)|\leq \frac{\pi}{2}-\delta,\qquad  \zeta\to\infty ,
 \end{equation}
for any  $0<\delta\leq \frac{\pi}{2}$, see also \cite{huba1998nrl}. The proof will be based on a generalized version of Watson's Lemma \cite{https://doi.org/10.1112/plms/s2-17.1.116}. To this end, let us define the Laplace transform
\begin{equation}\label{LaplaceT}
\mathcal{L}[f](\zeta) = \int_0^\infty f(x) e^{-\zeta x}\, dx, \quad \zeta\in\mathbb{C},
\end{equation}
of an integrable function $f: [0,\infty)\to\mathbb{C}$.

\begin{lemma}\label{GWatson}
[Generalized Watson's Lemma]
Assume that \eqref{LaplaceT} exists for some $\zeta=\zeta_0\in\mathbb{C}$ and assume that $f$ admits an asymptotic expansion of the form
\begin{equation}
f(x) =\sum_{n=0}^N a_n x^{\beta_n-1} + o(x^{\beta_N-1}), \qquad x>0,\quad x\to 0,
\end{equation}
where $a_n\in\mathbb{C}$ and $\beta_n\in\mathbb{C}$ with $\Re\beta_0>0$ and $\Re\beta_n>\Re\beta_{n-1}$ for $1\leq n\leq N$.
Then $\mathcal{L}[f](\zeta)$ admits an asymptotic expansion of the form
\begin{equation}
\mathcal{L}[f](\zeta) =\sum_{n=0}^N a_n \Gamma(\beta_n)\zeta^{-\beta_n} +o(\zeta^{-\beta_N}),\quad v,\quad \zeta\to \infty,
\end{equation}
for any real number $0<\delta\leq \frac{\pi}{2}$, where $\Gamma$ is the standard Gamma function. 
\end{lemma}
For a proof of the above Lemma, we refer e.g. to \cite{erdelyi1961general}. Classically, Lemma \eqref{GWatson} is applied to prove that the imaginary error function admits an asymptotic expansion for $x\in\mathbb{R}$ of the form
\begin{equation}\label{erfireal}
    \erfi(x)\sim \frac{e^{x^2}}{\sqrt{\pi}x}\sum_{k=0}^\infty \frac{(2k-1)!!}{(2x^2)^k},\qquad \text{ for } x>0,\quad x\to\infty,
\end{equation}
see also \cite{olver1997asymptotics}, based on the classical version of Watson's Lemma, whose assumptions are, however, unnecessarily restrictive \cite{wong1972generalization}. 

For completeness, we recall the derivation of \eqref{Ipsymptotic} based on Lemma \ref{GWatson}. First, let us rewrite $\erfi$ as a Laplace transform using the change of variables $t=\sqrt{1-s}$ with $dt=\frac{ds}{2\sqrt{1-s}}$
\begin{equation}
    \begin{split}
    \erfi(\zeta)&=\int_0^1\frac{d}{dt} \erfi(t\zeta)\,dt=\frac{2\zeta}{\sqrt{\pi}}\int_0^1  e^{t^2\zeta^2}\, dt = \frac{2\zeta}{\sqrt{\pi}}\int_0^1  e^{\zeta^2(1-s)}\, \frac{ds}{2\sqrt{1-s}}\\
    &= \frac{\zeta e^{\zeta^2}}{\sqrt{\pi}}\int_0^1 \frac{1}{\sqrt{1-s}}  e^{-s\zeta^2}\, ds=\frac{\zeta e^{\zeta^2}}{\sqrt{\pi}}\int_0^{\infty} \frac{\chi_{[0,1]}(s)}{\sqrt{1-s}}  e^{-s\zeta^2}\, ds.
    \end{split}
\end{equation}
From the Taylor expansion of the Binomial function, we know that
\begin{equation}
    \frac{1}{\sqrt{1-s}}=\sum_{n=0}^{\infty}\binom{-1/2}{n} (-s)^n=\sum_{n=0}^{\infty} 4^{-n}\binom{2n}{n}s^n,
\end{equation}
which allows us to apply Lemma \eqref{GWatson} with $\beta_n=n+1$ and $a_n=4^{-n}\binom{2n}{n}$, thus leading to
\begin{equation}
\begin{split}
    \erfi(\zeta)&\sim \frac{\zeta e^{\zeta^2}}{\sqrt{\pi}}\sum_{n=0}^\infty 4^{-n}\binom{2n}{n}\Gamma(n+1) \zeta^{-2(n+1)}\\
    &\sim \frac{e^{\zeta^2}}{\sqrt{\pi}}\sum_{n=0}^\infty \frac{(2n)!}{4^{n}n!} \zeta^{-2n-1}\\
    &\sim \frac{e^{\zeta^2}}{\zeta\sqrt{\pi}}\sum_{n=0}^\infty \frac{(2n-1)!!}{(2\zeta)^{n}},
    \end{split}
\end{equation}
for $\zeta\to\infty$ and $|\arg(\zeta)|\leq \frac{\pi}{2}-\delta$, $0<\delta\leq \frac{\pi}{2}$. This is consistent with formula \eqref{erfireal} for the limit along the real line. Finally, we arrive at the following asymptotic expansion for $Z$:
\begin{equation}
Z_{+}(\zeta)\sim \ri\sqrt{\frac{\pi}{2}}e^{-\frac{\zeta^2}{2}}-\sum_{n=0}^\infty  \frac{(2n-1)!!}{\zeta^{2n+1}},  \qquad \text{ for } |\arg(\zeta)|\leq \frac{\pi}{2}-\delta,\qquad  \zeta\to\infty,
\end{equation}
 which is, of course, equivalent to
 \begin{equation}
 Z_{+}(\zeta) \sim -\sum_{n=0}^\infty \frac{(2n-1)!!}{\zeta^{2n+1}},  \qquad \text{ for }|\arg(\zeta)|\leq \frac{\pi}{2}-\delta,\qquad  \zeta\to\infty ,
 \end{equation}
since $|e^{-\zeta^2}|^2=e^{-2(x^2-y^2)}\to 0$ for $\Re{\zeta}=x\to \infty $.

\bibliographystyle{abbrv}
\bibliography{DynamicalSystems}

\end{document}